\documentclass[a4paper,11pt]{article}
\usepackage{jheppub} 
\usepackage{lineno}
\usepackage{xcolor}
\usepackage{braket}
\usepackage{orcidlink}
\usepackage{slashed}
\usepackage{orcidlink}
\newcommand{\beq}{\begin{equation}}
\newcommand{\eeq}{\end{equation}}
\newcommand{\bea}{\begin{eqnarray}}
\newcommand{\eea}{\end{eqnarray}}
\newcommand{\mec}

\title{\boldmath The bearable inhomogeneity of the baryon asymmetry }

\author[a]{Hengameh Bagherian\orcidlink{0000-0001-6504-5187},}
\emailAdd{hengameh@g.harvard.edu}
\author[b,c]{Majid Ekhterachian\orcidlink{0000-0002-1980-2498},}
\emailAdd{majid.ekhterachian@sns.it}
\author[b]{Stefan Stelzl\orcidlink{0000-0001-5964-1054}}
\emailAdd{stefan.stelzl@epfl.ch}
\affiliation[a]{Department of Physics, Harvard University, Cambridge, MA 02138, U.S.A.}
\affiliation[b]{Theoretical Particle Physics Laboratory (LPTP),
Institute of Physics, EPFL, Lausanne, Switzerland}
\affiliation[c]{Scuola Normale Superiore, Piazza dei Cavalieri 7, 56126, Pisa, Italy}

\abstract{
We study the implications of precision measurements of light-element abundances, in combination with the Cosmic Microwave Background, for scenarios of physics beyond the Standard Model that generate large inhomogeneities in the baryon-to-photon ratio. We show that precision Big Bang Nucleosynthesis (BBN) places strong constraints on any mechanism that produces large-scale inhomogeneities at temperatures around or below the TeV scale. In particular, we find that fluctuations of order $25\%$ on comoving length scales larger than the horizon at $T \simeq 3~\mathrm{TeV}$ are incompatible with the observed light-element abundances.
This sensitivity to early-universe physics arises because baryon-number inhomogeneities homogenize primarily through diffusion, a slow process. As a result, BBN serves as a novel probe of baryogenesis below the TeV scale, readily ruling out some proposed scenarios in the literature. We discuss the implications for electroweak baryogenesis, and further show that precision BBN provides a new probe of first-order phase transitions that generate gravitational waves in the pHz–mHz frequency range. This yields constraints on the electroweak phase transition, as well as first-order phase transitions that have been suggested as an explanation of the pulsar timing array signal. Finally, we comment on the future prospects for improving this probe.}

\begin{document}
\maketitle
\flushbottom

\section{Introduction}
\label{sec:intro}

In the era of precision cosmology,  it is of great importance to leverage the available cosmological data as a new probe of physics beyond the standard model (BSM).
The success of Big Bang nucleosynthesis (BBN) in accurately predicting the observed abundances of light elements makes it one of the most powerful and precise probes of the early universe.
By combining measurements of primordial light element abundances with our understanding of BBN physics and the measured nuclear reaction rates as input, the baryon-to-photon ratio $\eta$ can be determined.
Independently, a measurement of $\eta$ to sub-percent precision is obtained from the angular power spectrum of the cosmic microwave background (CMB)~\cite{Planck:2015fie, Planck:2018vyg}.
The most precise BBN-based determination, dominated by the deuterium-to-hydrogen abundance ratio ($D/H$), achieves approximately $2\%$ precision~\cite{Yeh:2022heq} and is consistent with the value extracted from the CMB. This close agreement between BBN and CMB determinations of $\eta$ represents a significant success of modern cosmology.
Currently, the dominant uncertainties in the prediction of the baryon-to-photon ratio in the BBN stem from that of nuclear reaction rates, making recent experimental advances, such as those of the LUNA collaboration~\cite{Mossa:2020gjc}, particularly important to improve the uncertainty in the BBN predictions.

The CMB is a powerful probe of early-universe inhomogeneities, in particular probing baryon isocurvature perturbations at the level of $O(10^{-5})$~\cite{Planck:2018jri}. 
 In this work, we examine how BBN similarly probes baryon isocurvature perturbations, i.e., inhomogeneities in the baryon-to-photon ratio; for the effect of adiabatic modes on BBN, see e.g.~\cite{Inomata:2016uip,Jeong:2014gna}. 
The sensitivities of CMB and BBN to spatial inhomogeneities in $\eta$ differ significantly, offering complementary insights into potential physics beyond the standard cosmological model. 
In particular, while the level of precision from BBN is much less than that of the CMB, it probes length scales much smaller than those probed by the CMB. 
As we will show, BBN can be sensitive to comoving length scales that were sub-horizon even at electroweak temperatures, while CMB is primarily sensitive at comoving length scales of the order of and larger than the Hubble length at the time of recombination (see~\cite{Planck:2018jri} and e.g.~\cite{Buckley:2025zgh} for a recent study).

The deuterium-to-hydrogen ratio ($D/H$), being the most sensitive BBN probe for determining $\eta$, is particularly interesting in the context of sensitivity to inhomogeneities. Crucially, the final deuterium abundance depends non-linearly on the local value of $\eta$. Because of this non-linear dependence, two different spatial distributions of $\eta$ with the same average value generally yield distinct predictions for the $D/H$ ratio.
That is, averaging deuterium and hydrogen abundances across regions with varying baryon densities does not eliminate the effects of small-scale fluctuations in $\eta$. For small fluctuations around its mean value, the final abundances can be expanded in terms of the relative size of these fluctuations. While the linear term averages out, higher-order terms yield predictions different from homogeneous BBN. As a rough estimate, assuming an $O(1)$ coefficient for the second-order terms, a percent-level measurement of $D/H$ can probe fluctuations in $\eta$ of order $10\%$. This sensitivity applies to fluctuations on all scales that persist until BBN, including those significantly smaller than the horizon size at that time.

A remarkable observation emphasized in this work is that inhomogeneities introduced at times much earlier than BBN could survive until then and thus be probed. This applies not only to superhorizon fluctuations but also to fluctuations with length scales well within the horizon at the time of nucleosynthesis.
To understand the reach of BBN as a probe of early times, we must therefore consider the evolution of sub-horizon inhomogeneities in $\eta$. For small fluctuations, diffusion is the dominant mechanism for homogenizing baryons. As the temperature $T$ decreases in a radiation-dominated universe, the mean free path of particles (and thus the diffusion constants) grows, causing the diffusion timescale—set by the inverse Hubble scale $H^{-1}$—to increase. Consequently, the effects of diffusion are dominated by the latest times. The available time for diffusion is $t \sim H^{-1} \sim M_\text{Pl} / T^2$, whereas the typical interaction time and length scale for a strongly coupled system are $\tau \sim l_{\text{mfp}} \sim 1/T$. A particle thus undergoes approximately $N \sim t / \tau$ scatterings, and its random walk results in a diffusion distance $d \sim \sqrt{N}\,l_{\text{mfp}}$. Comparing this diffusion distance to the Hubble length, $l_H \sim H^{-1}$, yields $d / l_H \sim \sqrt{T/M_\text{Pl}} \ll 1$. This indicates that diffusion is inefficient at erasing inhomogeneities unless these fluctuations occur on length scales many orders of magnitude smaller than the horizon scale at late times.

It is instructive to estimate the earliest time at which sub-horizon inhomogeneities can be produced and still survive until BBN. Consider perturbations generated at the horizon scale at some temperature $T_i$. Due to cosmic expansion, the length scale of these inhomogeneities grows by the factor $(a_{\text{BBN}}/a_i) \sim (T_i/T_{\text{BBN}})$ by the time of BBN, becoming $(T_i/T_{\text{BBN}}) \cdot l_H(T_i)$, where the $a$ represents the corresponding scale factor. Comparing this length with the diffusion length at BBN, $d \sim \sqrt{T_{\text{BBN}}/M_\text{Pl}} \cdot l_H(T_{\text{BBN}})$, we see that fluctuations survive provided $T_i \lesssim \sqrt{T_{\text{BBN}} M_\text{Pl}} \sim 10^4\,\text{TeV}$. This naive estimate assumes strong coupling of baryons to the plasma; however, at the relevant temperatures, neutrons are only weakly coupled and dominate baryon diffusion. A more refined treatment (see section~\ref{sec:Diffusion}) yields the condition $T_i \lesssim 3~\mathrm{TeV}$. Thus, horizon-scale inhomogeneities generated at temperatures around a few $\mathrm{TeV}$ can just barely evade diffusion and survive until BBN, offering a tantalizing window into high-energy physics of the early universe.

We find that inhomogeneities greater than $\mathcal{O}(20-30\%)$ that are generated over distances larger than $\mathcal{O}(H^{-1})$ at temperatures of a few $\mathrm{TeV}$ are incompatible with the measured light element abundances. This novel constraint on the early universe has exciting implications, some of which we briefly explore. It readily rules out a recently proposed baryogenesis mechanism~\cite{Elor:2024cea}, which produces baryons with large spatial inhomogeneities, provides a test for electroweak baryogenesis (EWBG) scenarios involving domain walls~\cite{Azzola:2024pzq}, and perhaps most importantly, serves as a new probe of standard electroweak baryogenesis. Additionally, it constrains any model that imprints inhomogeneities at relevant length scales onto a pre-existing baryon (or equivalently $B-L$) asymmetry, such as scenarios involving strong first-order phase transitions in the early universe.

The first studies on the evolution of inhomogeneities and their effect on BBN were performed already in the 1980s~\cite{Applegate:1985qt,Applegate:1987hm}, followed by subsequent refinements in the treatment of diffusion~\cite{1990ApJ...358...36M,Kurki-Suonio:1992knt,Suh:1998nt}. In~\cite{Jedamzik:1993tcf,Jedamzik:1993dc}, the authors tracked baryon inhomogeneities from early times until BBN, taking into account not only baryon diffusion—the dominant process for small fluctuations—but also heat conduction via neutrinos (see also~\cite{Heckler:1993nc}, which found neutrino heat conduction relevant only for very large baryon overdensities).\footnote{Note that recently also the diffusion of light elements between CMB and reionization was studied \cite{Scherrer:2021tbo}.}  In the absence of concordance between CMB and BBN data at the time, as well as within BBN data itself, early studies also investigated whether inhomogeneities during BBN could resolve these discrepancies~\cite{Jedamzik:2001qc,Kainulainen:1998vh}. The possibility of constraining electroweak baryogenesis was previously discussed in~\cite{Fuller:1993sp,Heckler:1994uu,Megevand:2004ry,Brandenberger:1994fe}. Notably, during these earlier investigations, the baryon-to-photon ratio $\eta$ was still treated as a free parameter due to the absence of precise CMB measurements. Its value was determined later through precision CMB observations, most notably by the Wilkinson Microwave Anisotropy Probe (WMAP)~\cite{WMAP:2003ivt}, and subsequently refined by the Planck satellite~\cite{Planck:2018vyg}. 

More recently, the effect of inhomogeneities on BBN was reconsidered in~\cite{Barrow:2018yyg,Inomata:2018htm}. The analyses in these works were limited to length scales larger than the neutron diffusion length scale at BBN, so that diffusion until BBN could be neglected. In our analysis, we also consider length scales smaller than the neutron diffusion length at BBN. This is particularly important, since these small length scales are the ones that probe the universe at its earliest times.
Our results differ from those of~\cite{Barrow:2018yyg,Inomata:2018htm}, even on length scales where neutron diffusion can be ignored. The difference arises because, in determining the $D/H$ ratio, we separately average over deuterium and hydrogen densities.
Additionally, in our study of baryon diffusion we improve upon previous studies by solving the coupled Boltzmann equations for neutrons and protons including diffusion. This is particularly significant for accurately determining the evolution of proton inhomogeneities.

This work is organized as follows: In section~\ref{sec:Diffusion} we explain the big picture and
follow the inhomogeneities in the baryon-to-photon ratio from the early times until BBN by solving the coupled diffusion equations of protons and neutrons. In section~\ref{sec:BBNepoch}, we first review BBN and then derive the allowed magnitude of the inhomogeneities during BBN. For this purpose, we use the \texttt{PRyMordial} package~\cite{Burns:2023sgx} to obtain the dependence of the light element abundances on local proton and neutron densities. Finally in section~\ref{sec:boundingbaryogensis} 
we study various scenarios that produce inhomogeneities in the baryon-to-photon ratio and are constrained by our work. This includes models of baryogenesis and scenarios which source a large gravitational wave signal.

\section{Homogenization of baryons by diffusion}\label{sec:Diffusion}

In this section, we study the evolution of inhomogeneities in the baryon asymmetry from early times up to and including BBN. For small-amplitude inhomogeneities, the evolution is governed by baryon diffusion.

\subsection{Estimates and big picture} \label{sec:diffusionEstimate}
In this subsection, we give a first estimate of the diffusion length of baryons and the evolution of inhomogeneities in baryon number. A more rigorous treatment will be presented in the next subsection.

We start by writing the FRW metric as
\begin{equation}
    ds^2 = dt^2 - a^2 d\vec{x}^2\,,
\end{equation}
where $a$ is the scale factor and $\vec{x}$ denotes the comoving spatial coordinates. The diffusion equation for a comoving charge density \( n \) is given by 
\begin{align}\label{eq:diffone}
  \frac{\partial n}{\partial t} = D \nabla^2 n\,,
\end{align}
with $D$ being the diffusion coefficient. Here, we define $\nabla^2 = \nabla_i \nabla^i = \sum_i a^{-2}\partial_i \partial_i$. For simplicity, in this subsection we assume the relation $a(t) \propto T^{-1}$, neglecting modifications arising when different species (e.g., electrons) become nonrelativistic. With this assumption, we can express the Hubble scale as
\begin{equation}
H \equiv \frac{\dot{a}}{a} = -\frac{1}{T} \frac{\partial T}{\partial t}\,,
\end{equation}
which we can use to rewrite equation~\eqref{eq:diffone} as
\begin{align}\label{eq:conformal_diffEq}
  -H T \frac{\partial n}{\partial T} = D \, \frac{T^2}{T_0^2} \, \partial_x^2 n\,. 
\end{align}
In the above, $T_0$ denotes a reference temperature at which we normalize $a(T_0)=1$. The Hubble parameter during radiation domination is 
\begin{align}
    H = \frac{\pi}{3} \frac{T^2}{M_{\text{Pl}}} \left( \frac{g_*(T)}{10} \right)^{1/2} \,,
\end{align}
where \( g_*(T) \) is the effective number of relativistic degrees of freedom and $M_{\text{Pl}} \simeq 2.4\cdot 10^{18} \, \mathrm{GeV}$ is the reduced Planck mass. For now, we approximate $g_*(T) \simeq 10.75$, while we will incorporate its full $T$ dependence in the next subsection.

We are interested in determining the evolution of inhomogeneities in the baryon number density. To this end, we solve the corresponding diffusion equations. 
Before neutrino decoupling, the rates converting protons to neutrons and vice versa are faster than the Hubble rate, so we can treat their diffusion together. 
Let us first consider temperatures $T \gtrsim 1 \, \rm{MeV}$. Suppose there is some initial distribution of the baryon number density at a temperature $T_i$, well above the $\mathrm{MeV}$ scale. The effect of diffusion on this distribution is generally dominated by the lowest temperatures in the range considered, since as $T$ drops: (1) the Hubble scale drops, allowing more time for diffusion; and (2) interaction rates decrease, leading to larger diffusion coefficients. In particular, the diffusion of baryon number via quark diffusion at temperatures above the QCD phase transition is negligible.

At temperatures around $T \sim \mathrm{MeV}$, baryon diffusion is dominated by scattering with electrons. The mean free path of neutrons is significantly larger than that of protons, since protons are electrically charged while neutrons, being electrically neutral, interact only via their magnetic dipole moments. As a result, nucleons diffuse significantly only during the periods when they exist as neutrons. After neutrino decoupling, we must separately track neutron and proton distributions: neutrons continue to diffuse, but proton-electron interactions remain sufficiently strong that proton diffusion between neutrino decoupling and BBN is negligible.

The diffusion equation can be solved in momentum space as 
\begin{align}
  \tilde{n}_k(T) = \tilde{n}_k(T_i) \exp\left[ -k^2 \int_{T}^{T_i} \mathrm{d}T \, \frac{T}{H T_0^2} D \right] \equiv \tilde{n}_k(T_i) \exp\left[-\frac{d^2 k^2}{2}\right]\,,
\end{align}
here $\tilde{n}_k$ denotes the Fourier transform of $n$ evaluated at comoving momentum $k$. Let us now estimate the neutron diffusion coefficient. The scattering of electrons (and positrons) via the neutron's magnetic dipole moment has a cross section of the parametric form 
\begin{align}
    \sigma_{ne} \sim \frac{\alpha^2 \kappa^2}{m_n^2} \,,
\end{align}
with $\kappa = -1.29$ being the anomalous magnetic moment of the neutron in units of the nuclear magneton, and $m_n$ being the neutron mass.
The diffusion coefficient is estimated as $D \sim v_n^2 \Delta t$, where $v_n$ is the neutron velocity and $\Delta t$ is the effective mean free time, defined as the timescale over which a neutron changes its direction. This leads to the estimate $D_n \sim (\sigma_{ne} \, \mathfrak{n}_e)^{-1}$ for relativistic electrons ($T > m_e$), and $D_n \sim \sqrt{T/m_e} (\sigma_{ne} \, \mathfrak{n}_e)^{-1}$ for non-relativistic electrons ($T < m_e$), where $\mathfrak{n}_e$ is the electron number density. Throughout the text, we use the symbol $\mathfrak{n}$ to denote number densities per proper volume, while we reserve $n$ for comoving number densities. Parametrically, we have 
\begin{align} \label{eq:estimateelectrondiffusion}
    D_n \sim \frac{m_n^2}{\alpha^2 \kappa^2 T^3} \times \begin{cases} 
    1 & T > m_e \\
    (T/m_e)^2 e^{m_e/T} & T < m_e
    \end{cases}\,.
\end{align}
We will obtain a more precise expression in the next subsection. This gives us an estimate for the proton diffusion length
\begin{align}\label{eq:protondiffest}
d_p = \left(-2 \int_{T_\text{dec}}^{T_i} \mathrm{d}T \, X_n \frac{T}{H T_0^2} D_n \right)^{1/2} \simeq 10^5 \, \text{cm}\,.
\end{align}
In deriving the above, we took $T_0 = 1,\mathrm{MeV}$, and $T_\text{dec} \simeq 0.7,\mathrm{MeV}$ as the neutrino decoupling temperature. The quantity \hbox{$X_n(T)$} is the fraction of time nucleons spend as neutrons, which in equilibrium is given by $X_n^\text{eq}(T)= 1 / (1 + e^{Q/T})$, where $Q = 1.29,\mathrm{MeV}$ denotes the neutron-proton mass difference. Evaluating $X_n$ at $T \sim T_{\rm dec}$ gives $X_n \simeq 1/6$. After $T_{\rm dec}$ and until the onset of BBN, proton diffusion is negligible, while neutrons continue diffusing. Thus, the neutron diffusion length, primarily determined by the latest times around the BBN epoch, can be estimated as
\begin{align}\label{eq:neutrondiffest}
d_n = \left( -2 \int_{T_\text{BBN}}^{T_\text{dec}} \mathrm{d}T \, \frac{T}{H T_0^2} D_n \right)^{1/2} \simeq 10^7 \, \text{cm}\,,
\end{align}
with $T_\text{BBN} \simeq 0.06 \, \mathrm{MeV}$. 

To build an intuition for these length scales, let us compare them with the comoving Hubble scale at $100 \, \mathrm{GeV}$,
\begin{equation}
    (aH)^{-1} \simeq 3 \cdot 10^5 \, \text{cm}\,.
\end{equation}
We remind the reader once more that we have set the scale factor $a = 1$ at $T_0 = 1\,\rm{MeV}$. We see that even if inhomogeneities are generated at horizon scales around electroweak temperatures, their length scale remains large enough that proton inhomogeneities do not diffuse away before BBN. This implies that BBN can probe inhomogeneities created as early as the electroweak phase transition!

While we have now provided an intuitive picture and simple estimates of how far protons and neutrons diffuse before BBN, in the next subsection we refine these estimates by solving the coupled system of neutrons and protons across the temperature range of neutrino decoupling. In doing so, we keep the correct relationship between the scale factor $a$ and temperature $T$ across electron annihilation, and, of course, include all $\mathcal{O}(1)$ factors in the diffusion coefficients.

\subsection{Diffusion equations and coefficients}
\label{sec:diffusionrefined}
In this subsection, we rigorously calculate the diffusion length scales for protons and neutrons. The Boltzmann equations for the comoving neutron and proton number densities are
\begin{align}
    \partial_t n_n &= D_n \nabla^2 n_n - \Gamma_{n\to p} n_n + \Gamma_{p\to n} n_p \,, \\
    \partial_t n_p &= D_p \nabla^2 n_p + \Gamma_{n\to p} n_n - \Gamma_{p\to n} n_p \,.
\end{align}
The rates in the above set of equations consist of several processes:
\begin{align}
    \Gamma_{n\to p} &= \gamma_{n e^+ \to p \bar{\nu}} + \gamma_{n \nu \to p e} + \gamma_{n \to p e \bar{\nu}} \,, \\
    \Gamma_{p\to n} &= \gamma_{p \bar{\nu} \to n e^+} + \gamma_{p e \to n \nu} + \gamma_{p e \bar{\nu} \to n} \,,
\end{align}
with $\gamma_{x\to y}$ describing the thermal averaged rate of the process $x\to y$.
Detailed balance relates the rates as
\begin{align}
    \gamma_{n \nu \to p e} &= \gamma_{p e \to n \nu} \exp\left(-Q/T\right) \,, \\
    \gamma_{n e^+ \to p \bar{\nu}} &= \gamma_{p \bar{\nu} \to n e^+} \exp\left(-Q/T\right) \,,
\end{align}
where $\gamma_{n \to p e \bar{\nu}}$ describes neutron decay with lifetime $\tau_n \simeq 878~\text{s}$, and the three-body rate $\gamma_{p e \bar{\nu} \to n}$ is negligible. The rates $\gamma_{n \nu \to p e}$ and $\gamma_{n e^+ \to p \bar{\nu}}$ can be found, e.g., in~\cite{Mukhanov:2003xs}.

Using entropy 
conservation,
\begin{equation}
    S \propto g_{*S}(T) T^3 a^3 = \text{const}\,,
\end{equation}
we find
\begin{equation}
    a(T) = \frac{T_0}{T} \left( \frac{g_{*S}(T_0)}{g_{*S}(T)} \right)^{1/3}\,,
\end{equation}
giving
\begin{align}
    \frac{\partial}{\partial t} = H T \left(1 + \frac{1}{3} \frac{T}{g_{*S}(T)} \frac{\partial g_{*S}(T)}{\partial T} \right)^{-1} \frac{\partial}{\partial T} \,.
\end{align}
Substituting into the Boltzmann equations, we obtain in momentum space
\begin{multline}\label{eq:neutrondiff2}
     \frac{H T}{1 - \frac{1}{3} \frac{T}{g_{*S}(T)} \frac{\partial g_{*S}(T)}{\partial T}} \frac{\partial \tilde{n}_k^{(n)}}{\partial T} =\\ \left( D_n k^2 \left(\frac{T}{T_0}\right)^2 \left( \frac{g_{*S}(T)}{g_{*S}(T_0)} \right)^{2/3} - \Gamma_{n\to p} \right) \tilde{n}_k^{(n)} + \Gamma_{p\to n} \tilde{n}_k^{(p)} \,,
\end{multline}
and 
\begin{multline}\label{eq:protondiff}
     \frac{H T}{1 - \frac{1}{3} \frac{T}{g_{*S}(T)} \frac{\partial g_{*S}(T)}{\partial T}} \frac{\partial \tilde{n}_k^{(p)}}{\partial T} =\\ \left( D_p k^2 \left(\frac{T}{T_0}\right)^2 \left( \frac{g_{*S}(T)}{g_{*S}(T_0)} \right)^{2/3} - \Gamma_{p\to n} \right) \tilde{n}_k^{(p)} + \Gamma_{n\to p} \tilde{n}_k^{(n)} \,,
\end{multline}
\normalsize where again we have normalized $a=1$ at a reference temperature \( T_0 \). 
Since the equations are linear in number densities, the initial and final distributions are simply related by a transfer function (i.e., the retarded Green’s functions of the above equations). Once the temperature dependence of the diffusion coefficients $D_n$ and $D_p$ is known, the coupled Boltzmann equations can be numerically integrated to obtain the characteristic diffusion length as well as the full transfer functions. More explicitly, given an initial distribution of baryon asymmetry at an initial temperature $T_i$ sufficiently above the MeV scale, the distribution at later times is given by
\beq
\tilde{n}_k^{(N)}(T)=\tilde{G}^{(N)}(k, T, T_i) \, \tilde{n}_k^{(B)}( T_i)\,.
\eeq
Here, the $\tilde{G}^{(N)}$ functions are the momentum-space Green’s functions describing the evolution of the density distributions, where $N$ represents either neutrons or protons, and $\tilde{n}_k^{(B)}$ denotes the Fourier transform of the comoving baryon number density $n_B$. We have used the fact that for $T_i \gg \mathrm{MeV}$, the initial partition of $n_B$ into $n_p$ and $n_n$ has negligible effect on the final densities, since their ratios quickly approach equilibrium. Furthermore, for such high $T_i$, the Green’s functions $\tilde{G}^{(N)}$ become insensitive to $T_i$, allowing us to omit it from their arguments in the remainder of the paper. We define $X_{N}(T) = \tilde{G}^{(N)}(k=0,T)$ so that, consistent with the earlier notation, $X_{N}(T)$ denotes the ratio of spatially averaged nucleon densities to the baryon density. 

\paragraph{Neutron diffusion coefficient}

For the temperatures relevant to our analysis, the leading process controlling neutron diffusion is scattering with electrons, with scattering off protons being subdominant. These combine according to
\begin{align}
    D_n^{-1} = D_{ne}^{-1} + D_{np}^{-1} \,.
\end{align}
The transport cross section for electron-neutron scattering is given by~\cite{Applegate:1987hm}
\begin{align}
    \sigma^t_{ne} = 3 \pi \frac{\alpha^2 \kappa^2}{m_n^2} \,,
\end{align}
where \( \kappa = -1.29 \) is the anomalous magnetic moment of the neutron. Using this cross section, the corresponding diffusion coefficient can be obtained by first determining the mobility $b$, which relates the drag force $F$ to the particle velocity $V$ via $F = b^{-1} V$, and then using the Einstein relation $D = b T$ to relate the diffusion coefficient to the mobility. Assuming a Maxwell-Boltzmann distribution for the electrons, the diffusion coefficient is given by~\cite{Applegate:1987hm}
\begin{align}
    D_{ne} = \frac{\pi}{16} \left( \frac{m_n}{m_e} \right)^2 \frac{1}{m_e (\alpha \kappa)^2} \frac{e^{1/x}}{x(1 + 3x + 3x^2)} 
    \approx 0.67 \frac{e^{1/x}}{x(1 + 3x + 3x^2)}~\text{cm} \,,
\end{align}
where \( x \equiv T / m_e \). While the Maxwell–Boltzmann approximation is not strictly valid at high temperatures, its use introduces an error of less than $3\%$ even in that regime~\cite{Applegate:1987hm}. This expression exhibits the expected parametric behavior in both temperature regimes discussed in equation~\eqref{eq:estimateelectrondiffusion} of the previous subsection.

We now discuss the contribution to the diffusion coefficient from neutron–proton scattering. The transport cross section for neutron-proton scattering is 
\begin{align}
    \sigma_{np} = \frac{\pi a_s^2}{a_s^2k^2+(1-\frac{1}{2}r_sa_sk^2)^2} + \frac{3\pi a_t^2}{a_s^2k^2+(1-\frac{1}{2}r_ta_tk^2)^2}\,.
\end{align}
Here, $k$ is the nucleon momentum in the center-of-mass frame, and $a_s = -23.71~\mathrm{fm}$, $r_s = 2.73~\mathrm{fm}$, $a_t = 5.432~\mathrm{fm}$, and $r_t = 1.749~\mathrm{fm}$ denote the scattering lengths $a_i$ and effective ranges $r_i$ for the isospin-singlet (s) and isospin-triplet (t) channels, respectively.
We use the corresponding diffusion coefficient obtained from ref.~\cite{Jedamzik:2001qc} and given by
\begin{align}\label{eq:dnp}
    D_{np} = 9.4\times 10^{-12}\left(\frac{T}{\mathrm{MeV}}\right)^{1/2}\frac{\mathrm{MeV}^3}{\mathfrak{n}_p}\times \frac{1}{I(a_1,b_1)+0.16I(a_2,b_2)}~\mathrm{cm}\,,
\end{align}
where 
\begin{align}
    I(a,b) = \frac{1}{2}\int_0^{\infty}dx\frac{x^2e^{-x}}{ax+(1-bx/2)^2}\,,
\end{align}
with $a_1 = 13.59(T/\mathrm{MeV})$, $b_1 = -1.56(T/\mathrm{MeV})$, $a_2 = 0.71(T/\mathrm{MeV})$, and $b_2 = 0.23(T/\mathrm{MeV})$.\footnote{
The diffusion coefficient in equation~\eqref{eq:dnp} was obtained using the Chapman–Enskog approximation, which is valid for scattering of hard spheres. To evaluate the robustness of this result, we follow~\cite{Pitaevskii1981Physical} and compute the diffusion coefficient under the assumption that there is a hierarchy between masses (formally treating $m_n \gg m_p$, while sending both to their physical value at the end). We find that across all relevant temperatures, the diffusion coefficients differ by at most $30\%$. Since the effect of $D_{np}$ is subdominant compared to that of $D_{ne}$ for $T\gtrsim 100~\mathrm{keV}$, this change has negligible effect on the proton diffusion, it might however change the neutron diffusion lengths at BBN by $O(1)$.} In the low temperature limit equation~\eqref{eq:dnp}
simplifies to 
\begin{align}
   D_{np}=\frac{3\sqrt{\pi}}{8}\frac{1}{\mathfrak{n}_p(\pi a_s^2 + 3\pi a_t^2)}\left(\frac{T}{m_N}\right)^{1/2}\,.
\end{align}

\paragraph{Proton diffusion coefficient}

As described above, protons remain in thermal equilibrium with neutrons until the time of neutrino decoupling, and their spatial homogenization is primarily governed by neutron diffusion during this period. The diffusion coefficient for protons is dominated by scattering with electrons. Following~\cite{Applegate:1987hm}, the proton diffusion coefficient is given by
\begin{align}
    D_p = \frac{3\pi}{8 \alpha^2 \Lambda(T)} \left( \frac{\hbar}{m_e} \right) \frac{x e^{1/x}}{1 + 2x + 2x^2}
    \approx \frac{8.5 \times 10^{-7}}{\Lambda(T)} \frac{x e^{1/x}}{1 + 2x + 2x^2}~\text{cm} \,.
\end{align}
The divergence of the Coulomb scattering cross section in the forward region is regulated due to screening of the charge or potential at distances exceeding the Debye screening length. This is captured in the above equation by the Coulomb logarithm $\Lambda = \ln(2 / \theta_0)$, where $\theta_0^2 = k_d^2 / k_{\rm th}^2$ represents the square of the effective minimum scattering angle. Here, $k_d^2 = 4\pi \alpha \mathfrak{n}_e / T$ is the square of the Debye screening wave-vector, where $\mathfrak{n}_e$ is the sum of electron and positron number densities. The thermal wave-vector is given by $k_{\rm th}^2 \sim 3 m_e T$ in the non-relativistic regime and $k_{\rm th}^2 \sim (3 T)^2$ in the relativistic regime. The Coulomb logarithm is approximately constant, $\Lambda \approx 3.5$, at temperatures well above the electron mass ($T \gg m_e$), and increases to $\Lambda = \mathcal{O}(10)$ near BBN temperatures~\cite{Applegate:1987hm}. 

\paragraph{Results of diffusion}

We solve the coupled equations~\eqref{eq:neutrondiff2} and~\eqref{eq:protondiff}, using initial conditions in which protons and neutrons share the same momentum dependence and equilibrium number densities. At high temperatures, weak interaction rates are sufficiently rapid that any deviation from these initial conditions is quickly erased, and the system rapidly returns to equilibrium.

\begin{figure}
\centering
\includegraphics[width=\columnwidth]{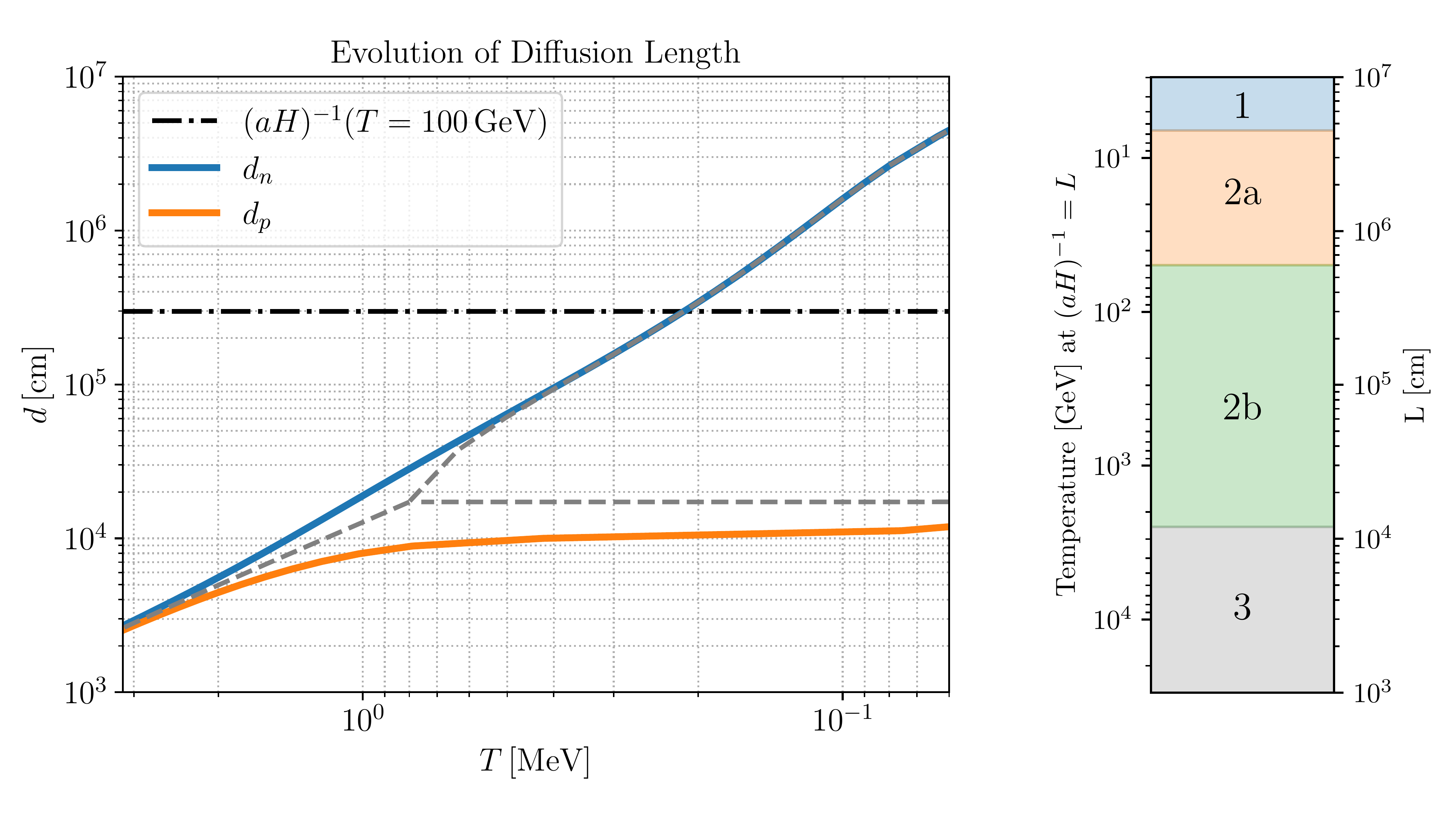}
\caption{ Evolution of the comoving diffusion length for protons and neutrons. The blue and yellow lines are obtained by solving the coupled eqs.~\eqref{eq:neutrondiff2} and~\eqref{eq:protondiff}. The dashed gray lines show results obtained by approximating the above equations using the instantaneous neutrino decoupling approximation. In this case we approximate weak rates interconverting protons and neutrons as efficient at $T>0.8~\mathrm{MeV}$ such that we can set their local ratios to the equilibrium value, while for $T<0.8~\mathrm{MeV}$ we treat these rates as negligible.
 For comparison, the Hubble horizon size at $100~\mathrm{GeV}$ is shown as the horizontal dot-dashed black line. 
On the right panel we show the different regimes relevant for BBN depending on the characteristic comoving length scale $L$ of the inhomogeneities. For the description of each regime, see the discussion at the end of section~\ref{sec:diffusionrefined}. Note that we have set the scale factor to unity at $T=1 \, \rm{MeV}$.}
\label{fig:dofT_trans}
\end{figure}

\begin{figure}
\centering
\includegraphics[width=0.7\columnwidth]{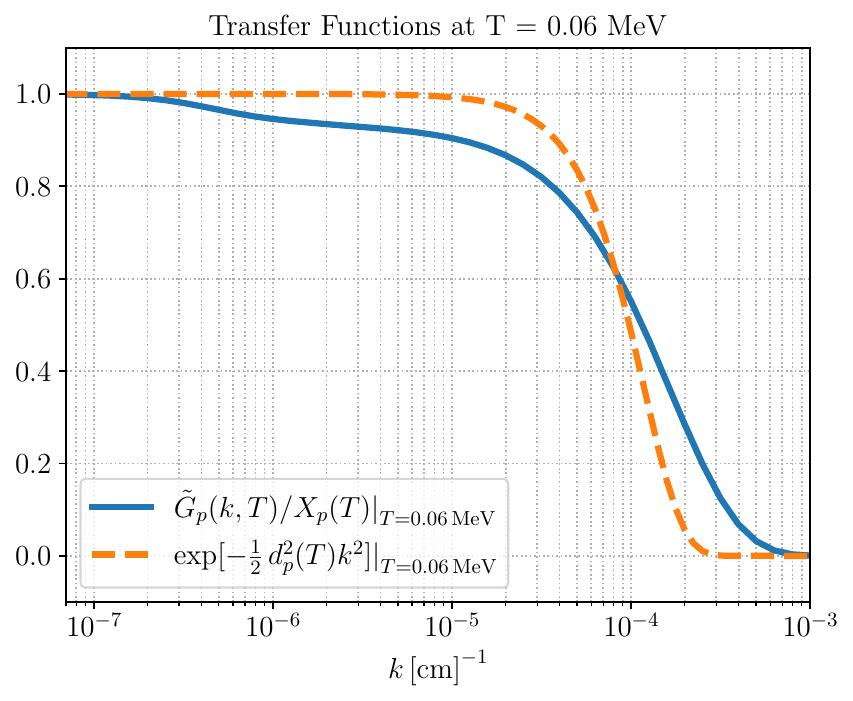}
\caption{The transfer function for the proton number density obtained by solving the coupled diffusion equations. For comparison a Gaussian profile, which would be obtained from a simple decoupled diffusion equation, is shown by fixing its width such that is has the same $d_p$.}
\label{fig:trans}
\end{figure}

Analogous to the case where the transfer function is a simple Gaussian, we define the diffusion length $d_i = 1/k_*^i$ via the momentum scale at which
\begin{equation}
   \frac{\tilde{G}^{(i)}(k_*^i,T)}{X_{i}(T)} = \exp[-1/2]\,,
\end{equation}
for $i = n, p$. The temperature evolution of these diffusion lengths is shown in figure~\ref{fig:dofT_trans}. These results can be compared with the naive estimates in eqs.~\eqref{eq:protondiffest} and~\eqref{eq:neutrondiffest}. While the refined treatment alters the diffusion lengths by $\mathcal{O}(1)$ factors, the overall conclusion remains unchanged: If inhomogeneities are generated on comoving length scales larger than $\sim 1/30 \, (a H)^{-1}(T = 100~\mathrm{GeV})$, they survive until BBN and may therefore be probed by the concordance between BBN and the CMB. To provide intuition for the significance of the diffusion length, in appendix~\ref{app:toymodelinhomogeneities} we introduce a toy model describing the initial baryon inhomogeneities and track their subsequent evolution under diffusion.
The transfer function for the neutron, $\tilde{G}^{(n)}$, is still well approximated by a Gaussian function, since neutron diffusion is dominated by late times long after neutrino decoupling. However, the transfer function for protons $\tilde{G}^{(p)}$ being sensitive to the time of decoupling, deviates significantly from a Gaussian.
In figure~\ref{fig:trans}, we show the transfer function for the proton at $T = 0.06~\mathrm{MeV} \simeq T_\text{BBN}$ and it is compared with a Gaussian transfer function with the same diffusion length.
 
Depending on the comoving length scale $L$ that characterizes the inhomogeneities, the system falls into one of the following regimes:
(1) For inhomogeneities on length scales larger than the neutron diffusion length, $L \gg d_n$, both neutron and proton inhomogeneities stay correlated and one can study the effect of inhomogeneities simply by varying $\eta$ and tracking the light element abundances as a function of its local value. (2) For inhomogeneities on length scales smaller than the neutron diffusion length and larger than the proton diffusion length, $d_p \ll L \ll d_n$, the inhomogeneities in protons are not altered, but neutrons are homogenized before BBN. This regime further splits into two distinct cases: (2.a) If neutron diffusion is slow compared to the rate at which neutrons are consumed to form deuterium, the neutron diffusion during BBN itself can be ignored, i.e.,
\beq
D_n (L a)^{-2} \ll \langle \sigma_{n p \rightarrow D \gamma } \, v \rangle n_p a^{-3}\,,
\eeq
where $\sigma_{n p \rightarrow D \gamma }$ is the cross section for deuterium production, $v$ is the relative velocity, and $\langle \cdots \rangle$ denotes the thermal average. This is the case for inhomogeneities with length scales larger than $\sim 6 \, \rm{km}$ at $\rm{MeV}$ temperature.
In this case, the effect of inhomogeneities can be studied by considering the dependence of the abundances on the local proton density and a constant average neutron density. (2.b) For inhomogeneities that are on smaller length scales, but still larger than $d_p$, neutron diffusion during BBN is also important. In the more proton-rich regions, more neutrons are consumed as deuterium is formed, leading to a larger reduction of the neutron density, which subsequently draws more neutrons from the neighboring regions to keep the neutron density homogeneous. The details of our analysis for this regime are given in appendix~\ref{app:neutronDBNN}, while the main text , section~\ref{sec:neutronBBNmain}, presents only the results. (3) Finally, if the length scale of inhomogeneities is smaller than the proton diffusion length, they are washed out before BBN. For example, from the transfer function shown in figure~\ref{fig:trans}, we see that at length scales around $5 d_p$, the amplitude of inhomogeneities is reduced by an order of magnitude. These distinct regimes are illustrated in the right panel of figure~\ref{fig:dofT_trans}. It is worth noting that we also considered and estimated the potential effect of proton diffusion during BBN at temperatures below $20\,\mathrm{keV}$, with the results presented in appendix~\ref{app:protonDBBN}..

\section{The tolerable inhomogeneity at BBN} \label{sec:BBNepoch}

In this section, we first review analytical estimates for the abundances of light elements, in particular ${}^4\mathrm{He}$ and $D$, and demonstrate why the $D/H$ ratio is especially sensitive to the value of $\eta$. This also gives an initial indication that the $D/H$ abundance depends non-linearly on both the baryon-to-photon ratio and the proton-to-neutron ratio. We then numerically evaluate the impact of inhomogeneities on the $D/H$ ratio using the \texttt{PRyMordial} package~\cite{Burns:2023sgx}. 
Our analysis leads to bounds on the allowed amplitude of inhomogeneities in the two regimes described in the previous section: (1) where both neutrons and protons stay inhomogeneous and (2) where neutrons homogenize but protons do not. We discuss the regimes (1), (2.a), and (2.b) in three different subsections, but leave a more detailed treatment of the case (2.b) to the appendix~\ref{app:neutronDBNN}. 
We conclude the section by commenting on possible future improvements to these bounds.

\subsection{A quick review of BBN}\label{sec:reviewBBN}

We briefly review the production of ${}^4\mathrm{He}$ and $D$ nuclei. Among light nuclei, ${}^4\mathrm{He}$ is the most tightly bound, and essentially all surviving neutrons become incorporated into it. This is because reaction rates forming heavier nuclei remain small at the time ${}^4\mathrm{He}$ is synthesized. Consequently, estimating the ${}^4\mathrm{He}$ abundance largely reduces to determining the neutron abundance at the onset of BBN.

BBN begins when significant amount of deuterium has formed. In equilibrium, the deuterium-to-proton ratio is governed by the binding energy \( B_D \simeq 2.2~\mathrm{MeV} \) and the baryon-to-photon ratio, 
\begin{equation}
    \left(\frac{n_D}{n_p}\right)_{\text{equil.}} \sim \mathfrak{n}_n \left(\frac{1}{m_p T}\right)^{3/2} e^{B_D/T} \propto \eta \left(\frac{T}{m_p}\right)^{3/2}e^{B_D/T} \,.
\end{equation}
Due to the small value of $\eta$, significant deuterium production occurs only at temperatures well below the binding energy. The suppression is overcome at \( T_{\text{BBN}} \simeq 60~\mathrm{keV} \).

To estimate the neutron fraction at this temperature, we note that as long as weak interactions remain in equilibrium, the neutron-to-proton ratio is given by
\begin{equation}
\frac{n_n}{n_p} \simeq e^{-Q/T} \,,
\end{equation}
where $Q \simeq 1.3~\mathrm{MeV}$ is the neutron–proton mass difference. These interactions freeze out around \( T_\nu \simeq 0.8~\mathrm{MeV} \), giving \( X_n(T_\nu) \simeq 1/6 \). From that point forward, only neutron decay alters the neutron abundance until BBN begins. Given the neutron lifetime \( \tau_n \approx 878~\text{s} \), this leads to
\begin{equation}
    X_n(T_{\text{BBN}}) \simeq 1/8 \,.
\end{equation}
This simple estimate gives the mass fraction of ${}^4\mathrm{He}$ as
\begin{equation}\label{eq:heab}
    Y_P \equiv \frac{\rho(^4\text{He})}{\rho_b} \simeq 2X_n \simeq 0.25 \,,
\end{equation}
with $\rho({}^4\mathrm{He})$ and $\rho_b$ denoting the mass densities of ${}^4\mathrm{He}$ and baryons, respectively. The onset temperature of BBN depends only logarithmically on $\eta$, resulting in a weak dependence of $Y_P$ on $\eta$. 
Although $Y_P$ is measured to approximately $2\%$ precision, it constrains $\eta$ only at $\mathcal{O}(1)$. In principle, $Y_P$ is also sensitive to inhomogeneities in $\eta$, but only at a similar precision. As we now show, the $D/H$ ratio is a much more sensitive probe.

The final abundance of deuterium is set by its freeze-out. At that point, neutrons are no longer available to form new deuterium and the abundance decreases through three main reactions: \( D + D \to {}^3\mathrm{He} + n \), \( D + D \to {}^3\mathrm{H} + p \), and \( D + p \to {}^3\mathrm{He} + \gamma \). The last of these, although suppressed by the electromagnetic coupling $\alpha$, becomes competitive at low enough $D$ abundances. This is critical. If the third reaction were negligible, the freeze-out deuterium abundance would be governed solely by the rates of the first two reactions and the Hubble expansion rate, making it insensitive to the local proton density. In this scenario, the final deuterium-to-hydrogen ratio would depend on $\eta$ only through the average proton density, scaling as $\langle D \rangle / \langle H \rangle \sim 1 / \langle \eta \rangle$. Thus, the $D/H$ ratio would be insensitive to inhomogeneities.

On the other hand, if the third reaction dominates, the deuterium abundance becomes exponentially sensitive to the local proton density, and therefore to $\eta$. This would make $D/H$ highly sensitive to inhomogeneities in $\eta$. 
As shown in~\cite{Mukhanov:2003xs}, for \( \eta \ll 10^{-9} \), the $D + D$ reactions dominate, while for \( \eta \gg 10^{-9} \), the $D + p$ reaction dominates. At the observed value \( \eta \simeq 6 \cdot 10^{-10} \), both reactions compete, resulting in significant nonlinear dependence---and thus sensitivity to inhomogeneities in $\eta$. 
Because the competing rate depends on the proton density, the deuterium abundance becomes sensitive to spatial variations in the proton-to-neutron ratio. In the next subsection, we verify this expectation through a numerical study.

In this work, we adopt the viewpoint that the lithium problem arises from our incomplete understanding of astrophysical systems. An interesting question, however, is whether inhomogeneities during BBN could resolve this lithium discrepancy. For small fluctuations, it is easy to see that this is not the case: small fluctuations in $\eta$ after averaging always lead to a larger lithium abundance. For large amplitude fluctuations, the situation is more complicated. A naively promising scenario is the case when inhomogeneities are so large in amplitude that in some regions there are excess neutrons staying after helium formation. These neutrons can then diffuse to other regions, leading to a late time neutron injection. Scenarios of late time neutron injection have been discussed in the literature as a solution to the lithium problem~\cite{AlbornozVasquez:2012emy,Coc:2013eha}, however more recent studies show that solving the lithium problem always introduces strong tension with the measured deuterium abundance~\cite{Coc:2014gia}.

\subsection{Current constraints}
In this section, we study the constraints on baryon inhomogeneities from BBN, considering three limiting cases. First, we examine the case where inhomogeneities are generated on large scales, such that neutron diffusion is too slow to homogenize the distribution before BBN. In this regime, the problem can be analyzed purely in terms of spatial variations in the baryon-to-photon ratio, $\eta$. Next, we consider the scenario where neutrons have completely homogenized before BBN, but protons have not. Depending on the length scale of the inhomogeneities, two distinct regimes emerge: one in which neutron diffusion during BBN can be neglected, and another in which it plays a significant role. In appendix~\ref{app:neutronDBNN}, we show that the final deuterium abundance in the latter case, labeled (2.b) earlier, is identical to that in case (1), where proton and neutron inhomogeneities are correlated.

\subsubsection{Inhomogeneity in $\eta$}
\label{sec:inhometa}
\begin{figure}
\centering
\begin{minipage}[c]{0.49\textwidth}
\centering
\includegraphics[width=\columnwidth]{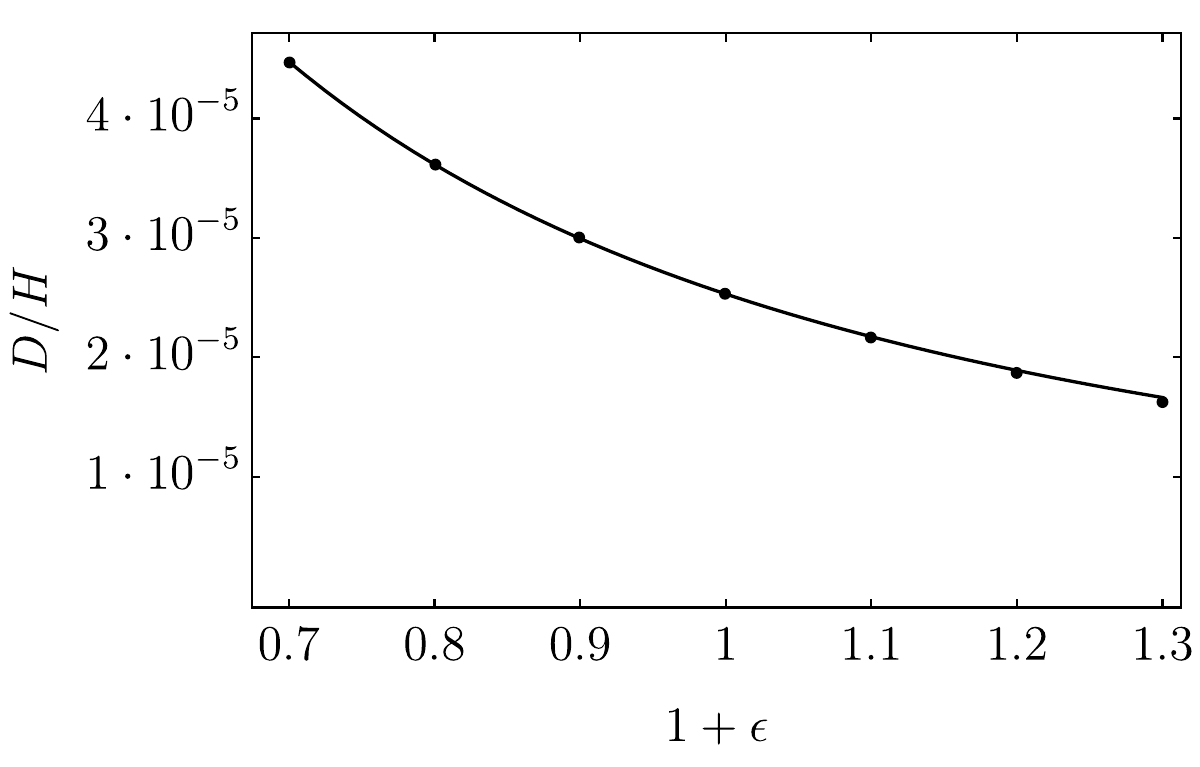}
\end{minipage}
\hfill
\begin{minipage}[c]{0.49\textwidth}
\centering
\includegraphics[width=\columnwidth]{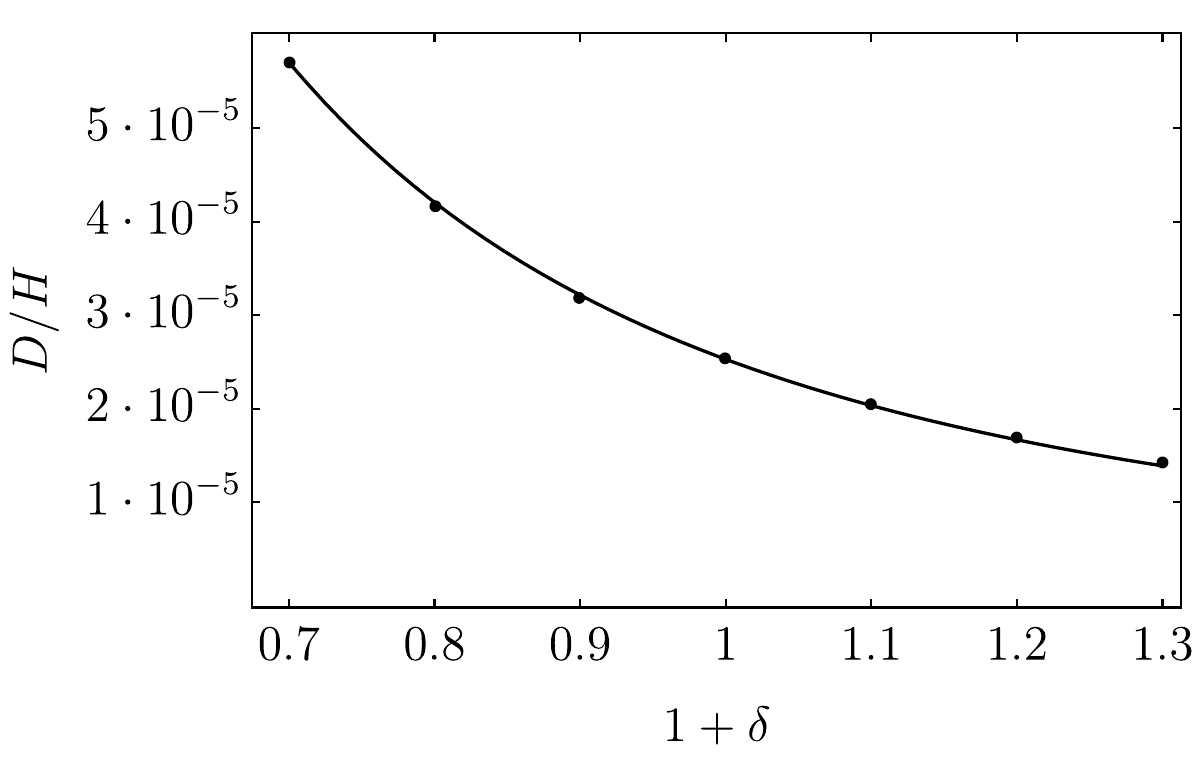}
\end{minipage}
\caption{\textbf{(Left)} Overlay of the numerical results with the fit formula from equation~\eqref{eq:etafit}, plotted against varying inhomogeneities in $\eta$.
\textbf{(Right)} Overlay of the fit formula from equation~\eqref{eq:fitbaryon} with its corresponding numerical data points. Here, the horizontal axis shows variations in proton density, while the neutron density is fixed to its standard BBN value.}
\label{fig:fitnumeric}
\end{figure}

We begin with the case where the inhomogeneity length scale is much larger than the neutron diffusion length. In this regime, proton and neutron densities share the same spatial dependence at the onset of BBN, captured by a spatially varying $\eta(x)$. We begin by studying this case in a region small enough to be characterized by a constant $\eta$, analyzing BBN as a function of $\eta$ and then considering the effects of spatial averaging. The CMB-inferred value, $\eta_{\text{CMB}}$, corresponds to the spatially averaged baryon-to-photon ratio, $\langle \eta \rangle$.
Assuming small-amplitude fluctuations, we parametrize $\eta(x)$ as $\eta_{\text{CMB}} (1 + \epsilon(x))$ and use a modified version of the \texttt{PRyMordial} package~\cite{Burns:2023sgx} to compute the $D/H$ abundance as a function of $\epsilon$ perturbation parameter. The results are well fit by
\begin{equation}\label{eq:etafit}
    \frac{D}{H} = 2.53 \cdot 10^{-5} \frac{1}{(1+\epsilon)^{1.6}}\,.
\end{equation}
The left panel of figure~\ref{fig:fitnumeric} shows the fit formula overlaid on our numerical results. We note that the power-law dependence $\eta^{-1.6}$ obtained from our fit agrees with the formula used in~\cite{Yeh:2022heq} to constrain $\eta$ from a combined BBN and CMB analysis, providing a consistency check of our methods.

To estimate the sensitivity to inhomogeneities, we first consider a simple case with two regions of equal size, each with $\eta = \eta_{\text{CMB}}(1 \pm \epsilon)$, and compute the spatial average. Since the hydrogen abundance scales linearly with $\eta$, we find
\begin{equation}
    \frac{\langle D \rangle}{\langle H \rangle} = 2.53 \cdot 10^{-5} \cdot \frac{1}{2} \left[ \left( \frac{1}{1+\epsilon} \right)^{0.6} + \left( \frac{1}{1-\epsilon} \right)^{0.6} \right] \simeq 2.53 \cdot 10^{-5} \left[ 1 + 0.48 \epsilon^2 + \dots \right]\,.
\end{equation}
For a more general $x$-dependence, a similar expression holds for small $\epsilon$, with the $\epsilon^2$ replaced by the variance $\langle \epsilon(x)^2 \rangle$, where angle brackets denote a spatial average.

Measurements of $D/H$ from quasar absorption lines have achieved percent-level precision (see~\cite{Cooke:2017cwo,Zavarygin:2017cov}). However, the dominant source of uncertainty arises from BBN predictions, primarily due to nuclear reaction rates. For concreteness, we adopt the conservative estimate from~\cite{Burns:2023sgx}:
\begin{equation}\label{eq:DHprecision}
\frac{D}{H} = (2.53 \pm 0.1) \times 10^{-5}\,,
\end{equation}
though more recent analyses (e.g.~\cite{Yeh:2022heq}) report slightly smaller uncertainties. This translates into a bound on the RMS amplitude of inhomogeneities, $\epsilon_{\text{RMS}} = \sqrt{\langle \epsilon(x)^2 \rangle}$:
\begin{equation} \label{eq:contraintonlyeta}
    \epsilon_{\text{RMS}} \le 0.28\,.
\end{equation}
Meaning, inhomogeneities in $\eta$ above $28\%$ are excluded by current BBN data.
\subsubsection{Inhomogeneities in protons only}

\label{sec:inhomprot}
Next, we consider the case where the characteristic length scale of inhomogeneities is much larger than the proton diffusion length but much smaller than that of neutrons. In this regime, protons remain inhomogeneous at the onset of BBN, while neutrons are fully homogenized.
As a result, both the neutron-to-proton ratio and the baryon-to-photon ratio vary spatially. As noted previously, we also consider the homogenization of protons during BBN and estimate the impact of proton diffusion in appendix~\ref{app:protonDBBN}. On sufficiently large scales, where proton diffusion remains insignificant above $20\,\mathrm{keV}$, this effect is safely negligible. On smaller scales, however, the results presented in this section can be modified by up to $15\%$.

As noted at the start of this section, even a uniform neutron density at the onset of BBN can develop inhomogeneities due to local nuclear reactions with spatially varying proton densities. The impact of these secondary inhomogeneities depends on the proton inhomogeneity scale, leading to two sub-regimes: one where neutron diffusion during BBN is negligible, and another where it efficiently smooths out these variations. In this subsection, we focus on the former—case (2.a) introduced above—while case (2.b), where diffusion is significant, is addressed below and in appendix~\ref{app:neutronDBNN}. 

Consider inhomogeneities in baryon number when neutrons and protons are in chemical equilibrium, $\eta(x) = \eta_{\text{CMB}} (1 + \epsilon(x))$. After they drop out of equilibrium, neutrons diffuse much further than protons, and homogenize before BBN. This leads to 
\begin{equation}
\label{eq:numberdensity2a}
    n_n(x,T) \propto X_n(T) \eta_\text{CMB} \qquad \text{and} \qquad n_p(x,T) \propto X_p(T) \eta_\text{CMB} (1+\epsilon(x))\,.
\end{equation}
Note that we neglect the small spatial dependence of $X_n$ and $X_p$. We evolve the \texttt{PRyMordial} code without modification until $T = 100~\mathrm{keV}$, by which point the neutron abundance is effectively frozen. At this temperature, we modify the inputs to satisfy equation~\eqref{eq:numberdensity2a}. As above, the calculation is performed in regions small enough that $\epsilon$ can be treated as constant.

Given that the code accepts $\eta$, $X_n$, and $X_p$ as inputs, the following expressions are used to ensure that equation~\eqref{eq:numberdensity2a} is satisfied: 
\begin{equation}\label{eq:epsilondeltarelation}
    \eta = \eta_{\text{CMB}} (1 + \delta) \qquad \text{with} \qquad \delta \equiv \epsilon \, X_p(T=100 \, \mathrm{keV})\,,
\end{equation}
and 
\begin{align}\label{eq:neutronfrac}
    X_n^{\text{mod}} &= \frac{n_n}{n_n + n_p} = \frac{X_n(T=100\mathrm{keV})}{1 + \delta}\,, \\
    X_p^{\text{mod}} &= \frac{n_p}{n_n + n_p} = \frac{X_p(T=100\mathrm{keV})}{1 + \delta} + \frac{\delta}{1 + \delta}\,, \label{eq:protonfrac}
\end{align}
where $X_n(T=100\mathrm{keV})$ and $X_p(T=100\mathrm{keV})$ are the standard neutron and proton fractions in a homogeneous universe evaluated at $100~$keV. By construction, these modified values still satisfy $X_n^{\text{mod}} + X_p^{\text{mod}} = 1\,$. With these modifications, the numerical results are well described by the fit
\begin{equation}\label{eq:fitbaryon}
D/H = 2.53 \cdot 10^{-5} \frac{1}{(1+\delta)^{2.28}}\,,
\end{equation}
where the numerical data points and the corresponding fit are shown in the right panel of figure~\ref{fig:fitnumeric}.

Neglecting neutron decay, the dependence of the final hydrogen number density on $\delta$ in a given patch can be approximated as
\begin{equation}
H \propto \left( X_p^{\text{mod}} - X_n^{\text{mod}} \right) \eta \propto \left( 1 + \frac{4}{3} \delta \right)\,,
\end{equation}
where this scaling follows from equations~\eqref{eq:epsilondeltarelation}, \eqref{eq:neutronfrac}, and~\eqref{eq:protonfrac}.
Averaging over two equal-sized patches with $\eta = \eta_{\text{CMB}} (1 \pm \delta)$, we obtain:
\begin{align}
    \frac{\langle D \rangle}{\langle H \rangle} & = 2.53 \cdot 10^{-5} \left\langle (1+\delta)^{-2.28} \left(1+\frac{4}{3} \delta\right) \right \rangle \simeq 2.53 \cdot 10^{-5} \left[1+ 0.7 \delta^2 + \dots  \right]\,,
\end{align}
where we have expanded for small $\delta$. Alternatively, a more careful approach using \texttt{PRyMordial} allows for separately averaging the deuterium and hydrogen abundances. This method properly accounts for neutron decay during the averaging procedure, yielding:
\begin{align}
    \frac{\langle D \rangle}{\langle H \rangle} & \simeq 2.53 \cdot 10^{-5} \left[1+ 0.8 \delta^2 + \dots  \right]\,,
\end{align}
Using equation~\eqref{eq:DHprecision}, this translates into a bound on the RMS amplitude of inhomogeneities, $\delta_{\text{RMS}} = \sqrt{\langle \delta(x)^2 \rangle}\,$:
\begin{equation}
    \delta_{\text{RMS}} \le 0.22\,,
\end{equation}
or equivalently,
\begin{equation} \label{eq:contraintonlyprotons}
   \epsilon_{\text{RMS}} \le 0.26\,.
\end{equation}

Thus, inhomogeneities that form on scales small enough for neutrons to homogenize by the time of BBN are constrained slightly more tightly than in scenarios where neutrons remain inhomogeneous. This aligns with the analytic understanding developed earlier—once neutrons are largely incorporated into ${}^4\mathrm{He}$, the remaining deuterium abundance at freeze-out becomes sensitive to the local proton number density. The increased sensitivity in the regime considered here can be understood in the following way: in the earlier case, where neutron and proton number densities were correlated at the onset of BBN, the inhomogeneity in the proton density at the time of deuterium freeze-out was reduced. This is because, in neutron-rich regions, more protons were consumed to form ${}^4\mathrm{He}$.

\subsubsection{Neutron diffusion during BBN}\label{sec:neutronBBNmain}
The final regime we consider, regime (2.b), corresponds to a scenario in which neutron diffusion remains active during BBN. This occurs when inhomogeneities form on the smallest scales that still allow proton inhomogeneities to persist until BBN. Recall that in this regime we have already assumed the inhomogeneities are on scales small enough for neutrons to fully homogenize before the onset of BBN. However, because BBN proceeds more rapidly in regions with higher proton density, this generates secondary neutron inhomogeneities, which are subsequently smoothed out by diffusion. We examine the case where diffusion during BBN is efficient, and demonstrate in appendix~\ref{app:neutronDBNN} that in this regime the proton number density following helium formation matches that in the scenario with inhomogeneities only in $\eta$ (regime 1). Given that the final deuterium abundance is primarily set by the remaining proton density (see section~\ref{sec:reviewBBN}), we arrive at the same bound as in section~\ref{sec:inhometa},
\begin{equation} \label{eq:boundregime2b}
   \epsilon_{\text{RMS}} \le 0.28\,.
\end{equation}

Intermediate cases, in which neutron diffusion during BBN is neither negligible nor fully efficient, fall between the regimes discussed in section~\ref{sec:inhometa} and section~\ref{sec:inhomprot}. A detailed numerical study of diffusion in this intermediate regime is left to future work.

Note that, up to this point, the definition of $\epsilon_\text{RMS}$ has implicitly assumed that the fluctuation length scale lies within a single regime. The more general case, in which inhomogeneities are initially generated across a range of length scales spanning multiple regimes, is discussed in the next section~\ref{sec:fluctuationdifferentregimes}. As expected, all modes that persist until BBN contribute to the modification of $D/H$, and the resulting constraint applies to a weighted combination of these modes, as shown in equation~\eqref{eq:differentscales}.

\subsection{Future prospects}

Significant improvements in both CMB observations and $D/H$ measurements are expected in the near future. The extraction of the baryon density $\Omega_b$, and thus the baryon-to-photon ratio $\eta$, is projected to improve by a factor of about three with the Simons Observatory~\cite{SimonsObservatory:2018koc}, and by an additional factor of two with CMB-S4~\cite{CMB-S4:2016ple}. Current $D/H$ measurements are statistically limited, and the forthcoming generation of extremely large telescope facilities with $\gtrsim30\,\mathrm{m}$ aperture will provide access to fainter and more distant quasars, thus improving the statistics by a factor of $\gtrsim 100$~\cite{Cooke:2024nqz}. Future facilities will also target higher-redshift systems, allowing cleaner measurements~\cite{Cooke:2024nqz}.

The primary limitation in constraining $\eta$ and its inhomogeneities from measurements of $D/H$ arises from uncertainties in the key nuclear reaction rates governing deuterium freeze-out. In particular, the reactions $D + p \to {}^3\mathrm{He} + \gamma$, $D + D \to n + {}^3\mathrm{He}$, and $D + D \to p + {}^3\mathrm{H}$ play a central role. Recent measurements by the LUNA collaboration~\cite{Mossa:2020gjc} have significantly improved the precision of the $D + p \to \gamma + {}^3\mathrm{He}$ reaction rate, rendering its contribution to the current $D/H$ error budget subdominant~\cite{Yeh:2020mgl}. The dominant uncertainties now arise from the two $D + D$ reactions, for which further improvements are anticipated in the near future. These advances are expected to enable a sub-percent level determination of $\eta$ from BBN~\cite{talkgustavino}. With the expected advances in CMB and light element observations, further improvements in all relevant nuclear rates will be essential. Assuming that the nuclear rates can keep up in precision with the observational advances, we show a projection for future improvement by a factor of $10$ in figure~\ref{fig:cmbVSbbn}.

\subsection{Constraint on inhomogeneities spanning over a wide range of scales}
\label{sec:fluctuationdifferentregimes}

In this subsection, we discuss constraints on inhomogeneities that extend over a broad range of length scales, rather than being confined to a single regime among $(1)$, $(2.a)$, $(2.b)$, or $(3)$.
Let the initial fluctuation in the baryon-to-photon ratio be parameterized as $\eta_i(x) = \eta_{\mathrm{CMB}}\big(1 + \epsilon_i(x)\big)$, and denote the Fourier transform of $\epsilon_i(x)$ by $\tilde{\epsilon}i(k)$. The relative fluctuation in the proton number density at the onset of BBN, $\tilde{\epsilon}_{\mathrm{BBN}}^{(p)}(k)$, is then given by
\beq \label{eq:evolvedeps1}
\tilde{\epsilon}_{\mathrm{BBN}}^{(p)}(k)= \frac{\tilde{G}^{(p)}(k,T_{\mathrm{BBN}})}{X_p(T_{\mathrm{BBN}})}\, \tilde{\epsilon}_i(k)\,.
\eeq
The initial variance of the fluctuations can be written in terms of a dimensionless power spectrum $\Delta^2(k)$, 
\beq
\langle \epsilon^2_i\rangle = \int d\ln k\, \Delta^2(k)\,.
\eeq
Our general constraints then follow from equation~\eqref{eq:evolvedeps1} and can be expressed as 
\small
\begin{align} \label{eq:differentscales}
   \int_{(1)} d\ln k\, \Delta^2(k)\frac{G(k)^2}{X_p^2}
    + 1.2\int_{(2a)} d\ln k\, \Delta^2(k)\frac{G(k)^2}{X_p^2}
    + \int_{(2b)} d\ln k\, \Delta^2(k)\frac{G(k)^2}{X_p^2} < 0.28^2\,,
\end{align}
\normalsize
where the subscripts $(1)$, $(2a)$, and $(2b)$ denote integration over the comoving momentum ranges corresponding to each of these regimes. The numerical factor $1.2\approx 0.28^2/0.26^2$ appears because proton number density fluctuations at deuterium freeze-out are larger in scenario $(2a)$, as discussed in section~\ref{sec:inhomprot}. This expression reduces to eqs.~\eqref{eq:contraintonlyeta}, ~\eqref{eq:contraintonlyprotons}, and~\eqref{eq:boundregime2b} when the power spectrum is restricted to length scales associated with the corresponding regimes.

\begin{figure}
\centering
\begin{minipage}[c]{0.49\textwidth}
\centering
\includegraphics[width=\columnwidth]{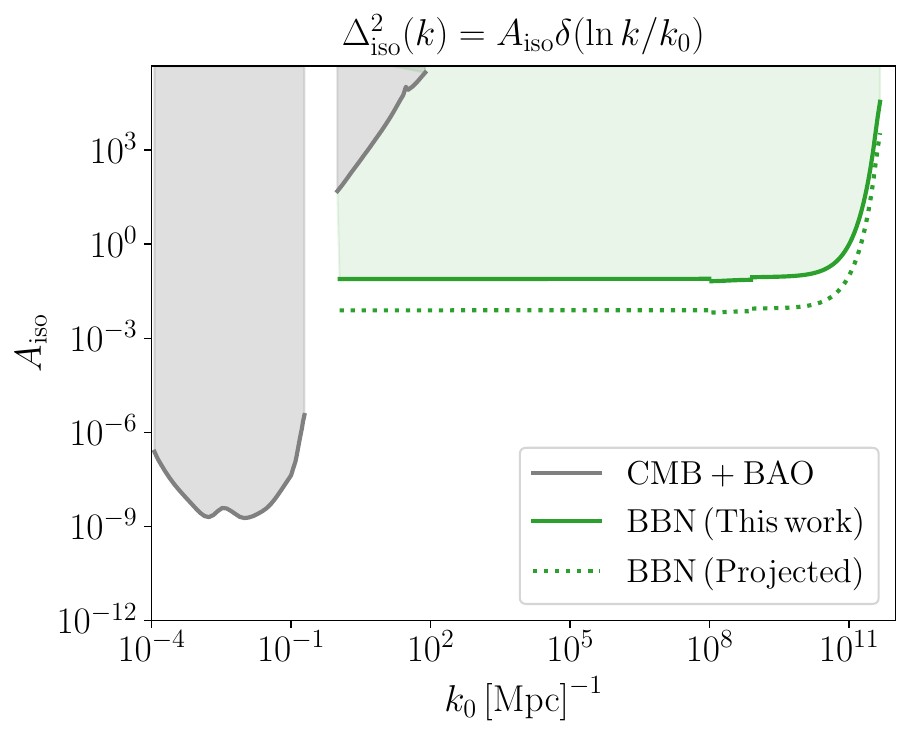}
\end{minipage}
\hfill
\begin{minipage}[c]{0.49\textwidth}
\centering
\includegraphics[width=\columnwidth]{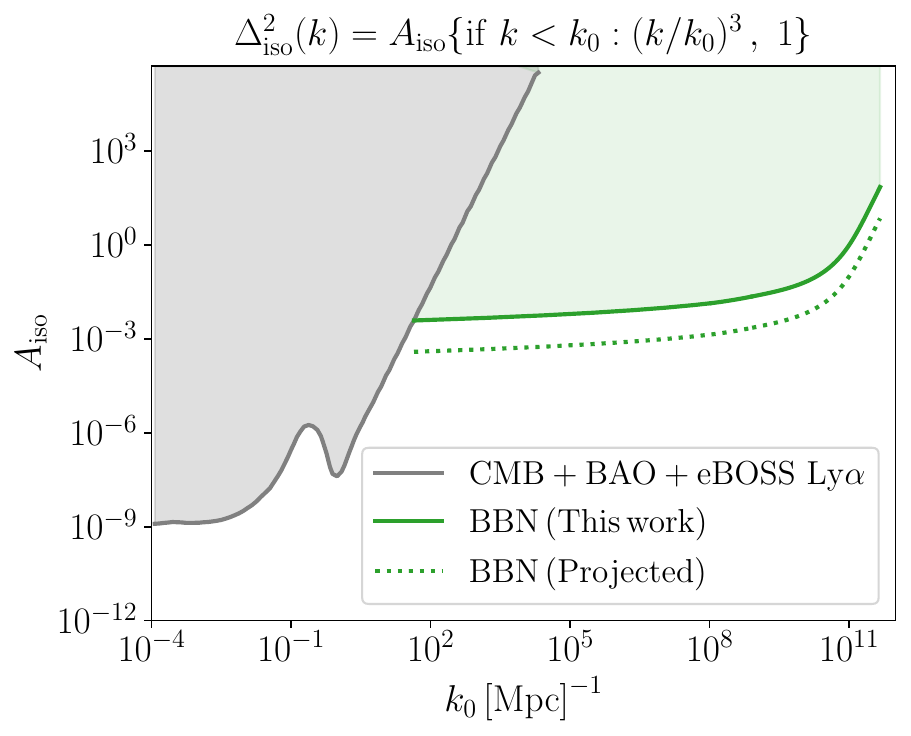}
\end{minipage}
\caption{Constraints on inhomogeneities that span a wide range of length scales, compared with existing bounds from~\cite{Buckley:2025zgh}. The dotted lines indicate the potential future reach, assuming a tenfold increase in sensitivity.
\textbf{(Left)} Constraints for a delta-function power spectrum, as defined in equation~\eqref{eq:deltapower}. Note the slightly stronger sensitivity of BBN to inhomogeneities in regime (2.a), consistent with equation~\eqref{eq:contraintonlyprotons} as discussed in the main text. The apparent gap between existing bounds from~\cite{Buckley:2025zgh} is largely dominated by Lyman-$\alpha$ forest data, which is why we did not extend our constraints into that region; they would be superseded by Lyman-$\alpha$. These bounds were not included in~\cite{Buckley:2025zgh} since they lie outside the validity of their approximations.
\textbf{(Right)} Same as left, but for a broken power law as defined in equation~\eqref{eq:brokenpower}.}\label{fig:cmbVSbbn}
\end{figure}

As an illustrative example, first consider a scale-invariant baryon number power spectrum, i.e., $\Delta^2(k)\equiv A_{\mathrm{iso}}$, for all length scales larger than the comoving horizon at BBN. In this case, the constraint on the dimensionless power spectrum becomes
\beq
A_{\mathrm{iso}} < 2.6\times 10^{-3}\,.
\eeq
In figure~\ref{fig:cmbVSbbn} we show the constraints for two different forms of the power spectrum, a monochromatic power spectrum
\beq\label{eq:deltapower}
\Delta^2(k)= A_{\rm iso} \delta\left(\ln(k/k_0)\right)\,,
\eeq
and a power spectrum that is scale invariant below some length scale and has a power-law suppression on small length scales,
\beq \label{eq:brokenpower}
\Delta^2(k)=  \begin{cases}
    A_{\rm iso}, \quad & \text{for} \qquad k>k_0\\
    A_{\rm iso} \left( \frac{k}{k_0}\right)^3 \quad & \text{for} \qquad k<k_0
\end{cases}\,.
\eeq
For comparison, we also show existing bounds taken from~\cite{Buckley:2025zgh}.

\section{Constraints on BSM scenarios that produce baryon inhomogeneities}\label{sec:boundingbaryogensis}

In this section, we examine the implications of our novel constraints for different scenarios that produce inhomogeneities in the baryon-to-photon ratio. We begin with baryogenesis mechanisms that generate an inhomogeneous baryon distribution, and then turn to models that imprint inhomogeneities onto an initially homogeneous universe.

\subsection{Constraints on baryogenesis mechanisms}

Models that generate baryons at temperatures below the $\mathrm{TeV}$ scale and produce sizable inhomogeneities in the baryon-to-photon ratio can be probed by BBN. In what follows, we first consider the important case of electroweak baryogenesis and then briefly comment on other recently proposed mechanisms that generate the baryon asymmetry inhomogeneously.

\subsubsection{Electroweak baryogenesis}

Arguably, the most well-motivated target to probe is electroweak baryogenesis, in which baryons are produced at the bubble walls during a first-order electroweak phase transition. The resulting baryon distribution is generically inhomogeneous for several reasons. First, key quantities that determine the baryon yield—such as the sphaleron rate and the velocity of the bubble wall—depend on temperature. As a result, as bubbles expand and the universe cools, the density of baryons they produce varies. Moreover, inhomogeneous reheating and bubble collisions can introduce additional spatial fluctuations in the baryon-to-photon ratio, particularly in scenarios with significant supercooling.

The timescale of a cosmological first-order phase transition is characterized by the parameter $\beta^{-1}$, as is standard in the literature (see e.g.\cite{Turner:1992tz}). The corresponding length scale of the resulting inhomogeneities is set by the typical bubble separation, $\sim v_w/\beta$, where $v_w \lesssim 1$ denotes the bubble wall velocity. Using the proton diffusion length derived in section~\ref{sec:diffusionrefined}, we find that for a transition occurring at $T \sim 100 \, \mathrm{GeV}$, and for $\beta/H \gtrsim \mathcal{O}(30)$, the bubble separation is sufficiently small that diffusion efficiently erases the inhomogeneities. Conversely, if $\beta/H \lesssim \mathcal{O}(30)$, the inhomogeneities can persist until the epoch of BBN, potentially leaving observable imprints.

We further expect the magnitude of inhomogeneities to be governed by the parameter $(\beta/H)^{-1}$ and to scale linearly with it when $(\beta/H)^{-1}$ is sufficiently small. The precise dependence on this parameter, however, can only be determined through numerical simulations. A detailed study of both the magnitude and spatial profile of the baryon fluctuations is left for future work~\cite{Futurework1}. Nonetheless, our results indicate that BBN may already constrain a significant portion of the electroweak baryogenesis parameter space, with improvement possible up to $\beta/H \sim \mathcal{O}(30)$. Remarkably, this range coincides with the regime of electroweak baryogenesis that produces the strongest gravitational wave signals.

\subsubsection{Other baryogenesis scenarios}

Any model that produces baryons inhomogeneously at large enough length scales is subject to the BBN constraint. In this section, we briefly discuss two additional baryogenesis scenarios that are in tension with the constraints presented in this work.

First, we comment on a scenario of baryogenesis via domain walls~\cite{Azzola:2024pzq}. In this scenario, the SM is extended by a singlet scalar field that acquires different expectation values in different domains, separated by domain walls. These walls rapidly enter the so-called scaling regime, leaving only a few domains per Hubble volume. Owing to the coupling of the singlet scalar to the SM Higgs, the Higgs vacuum expectation value (VEV) is suppressed within the core of the domain wall. Denoting the Higgs expectation value outside the wall by v and inside the core by $v_\text{core}$, one finds that at temperatures satisfying $v \gg T \gg v_\text{core}$, the sphaleron rate remains unsuppressed inside the wall but is exponentially suppressed outside. As the domain walls sweep through space, baryon asymmetry is generated at their location, in close analogy with electroweak baryogenesis. Because the walls move slowly, the resulting baryon production is expected to be highly inhomogeneous.\footnote{An explicit calculation of the baryon asymmetry in this model was not provided in~\cite{Azzola:2024pzq}. While this makes it difficult to quantitatively estimate the inhomogeneities, they are expected to be sizable.} These fluctuations are produced on length scales of $\mathcal{O}(H^{-1})$ at the electroweak epoch and therefore survive until the time of BBN. Consequently, this scenario appears incompatible with the observed light-element abundances.

Another baryogenesis scenario that conflicts with our constraints was recently proposed in~\cite{Elor:2024cea}. In this model, CP violation from B-meson oscillations within the SM serves as the source of CP violation needed to generate the baryon asymmetry. The remaining Sakharov conditions are realized as follows: B-mesons are produced through the out-of-equilibrium decay of a heavy particle into b-quarks, and these mesons subsequently decay into dark-sector states carrying baryon number, mediated by a new heavy particle. For SM CP violation to be sufficient, this mediator must be light in the early universe at $T \sim 100 \, \mathrm{MeV}$. However, collider bounds require the mediator to be heavy today, necessitating a temperature-dependent mass that transitions below $\mathcal{O}(100~\mathrm{MeV})$. This variation is implemented via a domain wall network, with the mediator light in some domains and heavy in others. As a result, significant baryon asymmetry is generated only in the domains where the mediator is light, producing large inhomogeneities on length scales that persist until BBN. This places the model in conflict with the observed abundances of light elements.

\subsection{Correlation with gravitational waves}
\label{sec:boundingGW}
Our novel constraints are not restricted to models in which baryons are generated inhomogeneously. Inhomogeneities can also be imprinted onto the baryon-to-photon ratio well after baryogenesis has taken place. Notably, models that predict a significant gravitational wave (GW) signal often generate baryon inhomogeneities at comparable length scales, thereby making them testable through BBN. As discussed in section~\ref{sec:Diffusion}, inhomogeneities with comoving length scales larger than the proton diffusion length at BBN survive until that epoch and are therefore constrained. Remarkably, the proton diffusion length at BBN corresponds today to a wavelength of $\mathcal{O}(10^9)\,\mathrm{km}$, or a frequency of $f \sim 0.1\,\mathrm{mHz}$. By contrast, the horizon size at BBN corresponds to a frequency of $f \sim \mathrm{pHz}$. One can therefore expect that cosmological sources active prior to BBN that produce a sizable GW background in this frequency range can also be probed by BBN. This includes, in particular, the frequency range accessible to pulsar timing arrays (PTAs) as well as part of the LISA sensitivity band.

An important source of a stochastic GW background is a first-order phase transition proceeding through the nucleation of bubbles. The GW signal from such a transition is maximized when the bubble separation is large and a significant fraction of the energy density is stored in the bubbles. Beyond generating GWs, a first-order phase transition also imprints inhomogeneities in the baryon-to-photon ratio, particularly in the case of strong supercooling. This arises because regions where bubbles nucleate later experience prolonged expansion, giving rise to adiabatic fluctuations. In addition, reheating proceeds inhomogeneously across space, generating what we refer to in this paper as non-adiabatic fluctuations, i.e., variations in the baryon-to-photon ratio. The evolution of both adiabatic and non-adiabatic modes is discussed in appendix~\ref{app:Tfluctuations}. We therefore expect a correlation between a strong GW signal and baryon inhomogeneities on large scales. Although we do not explore a possible first-order QCD phase transition in detail, we note that it would likely generate even larger inhomogeneities, since the properties of particles carrying baryon number change drastically across the two phases.

In this section, we first examine inhomogeneities induced by a strong first-order electroweak phase transition, which could be probed by LISA~\cite{LISA:2017pwj}, and then turn to two scenarios proposed to explain the recently observed PTA signal~\cite{Xu:2023wog,NANOGrav:2023gor,EPTA:2023fyk,Reardon:2023gzh}.

\subsubsection{LISA frequency range} \label{sec:lisa}

A GW signal from a first-order EW phase transition may be within the discovery potential of LISA~\cite{LISA:2017pwj}. For a phase transition occurring around $T \simeq 100 \rm{GeV}$, LISA is sensitive to $\beta/H \lesssim O(100)$ and the strength $\alpha \equiv \frac{\rho_\text{vac}}{\rho_\text{rad}} \gtrsim O(0.1)$ ~\cite{Caprini:2024hue}, where $\rho_\text{vac}$ is the energy density of the false vacuum and $\rho_\text{rad}$ the energy density of radiation during the phase transition.

Inhomogeneities are generated on length scales set by the typical bubble separation, which is of order $(H/\beta)\, l_H$. Thus, a phase transition with $\beta/H = \mathcal{O}(30)$ produces inhomogeneities that can persist until BBN. The magnitude of these inhomogeneities is expected to be governed by a combination of $\beta/H$ and $\alpha$, although a precise determination would require numerical simulations.

However, let us estimate the amplitude of inhomogeneities imprinted by a supercooled phase transition with $\alpha \gtrsim 1$. The duration of the transition, from the nucleation of the first bubbles until completion, is of order $\delta t \sim \mathcal{O}(\text{few})/\beta$. Since the bubbles that nucleate first also reheat first and subsequently cool through expansion, we estimate
\begin{equation}\label{eq:firstorderPTinhomo}
\frac{\Delta T}{T} \sim \frac{\mathcal{O}(\text{few})\, H}{\beta}\,,
\end{equation}
which initially does not produce inhomogeneities in the baryon density itself\footnote{There is another mechanism that generates inhomogeneities in the baryon density. Parts of the space in which bubbles nucleate later experience prolonged expansion, leading to to smaller baryon density and temperature in those regions. However, these fluctuations are adiabatic and don't contribute to the baryon-to-photon ratio. As discussed in appendix~\ref{app:Tfluctuations}, adiabatic modes that are generated at these length scales are damped and have no effect.}. Instead, this corresponds to inhomogeneities in the baryon-to-photon ratio of order
\begin{equation}\label{eq:foptinhomofinal}
\Delta \eta \simeq 3\frac{\Delta T}{T} \bar{\eta} \simeq \mathcal{O}(10)\, \frac{H}{\beta}\, \bar{\eta}\,,
\end{equation}
where we have used $\eta \propto T^{-3}$ together with equation~\eqref{eq:firstorderPTinhomo}. The temperature fluctuations in the fluid subsequently homogenize. As discussed in appendix~\ref{app:Tfluctuations}, for modes generated at the electroweak scale this homogenization is underdamped, so the baryon-to-photon inhomogeneities remain. We therefore expect equation~\eqref{eq:foptinhomofinal} to provide a good estimate of the final baryon density inhomogeneities after the fluid has relaxed, but prior to baryon number diffusion.

This highlights the potential of BBN to probe first-order phase transitions and underscores the need for a detailed study of the baryon-to-photon inhomogeneities imprinted in scenarios both with and without significant supercooling.

\subsubsection{Pulsar timing array signal}

The GW signal recently observed by PTAs~\cite{Xu:2023wog,NANOGrav:2023gor,EPTA:2023fyk,Reardon:2023gzh} could be explained by a cosmological phase transition, as discussed in~\cite{NANOGrav:2023hvm} and related works. Such a phase transition could take place both in a completely dark sector or in a sector coupled to the SM. Typical parameter values consistent with the PTA signal are $\alpha \sim \mathcal{O}(1)$ and $\beta/H \sim \mathcal{O}(10)$ at temperatures $T \sim \mathcal{O}(100~\mathrm{MeV})$.

If such a phase transition is coupled to the SM, it is expected to induce inhomogeneities in the baryon-to-photon ratio on length scales of order 1/(10H). These fluctuations are primarily driven by temperature variations arising from differences in reheating times (see section~\ref{sec:lisa}). As discussed in appendix~\ref{app:Tfluctuations}, the subsequent relaxation of temperature fluctuations in the fluid leads to two distinct regimes, depending on the mode length scale. Modes with $\lambda_\nu k_{\rm phys} \ll$ 1, where $\lambda_\nu$ denotes the neutrino mean free path, are underdamped and leave the baryon-to-photon inhomogeneities intact. By contrast, modes with $\lambda_\nu k_{\rm phys} \gtrsim 1$ are strongly damped, reducing the baryon-to-photon inhomogeneities. Nevertheless, the surviving inhomogeneities are likely to be in conflict with our constraints, although a dedicated study is required for a definitive assessment.

We now turn to the case in which the phase transition occurs purely within a dark sector. We first note that this scenario is already in tension with recent constraints on $\Delta N_{\rm eff}$\cite{Yeh:2022heq,ACT:2025tim}. Such a transition is expected to generate correlated fluctuations in temperature and baryon density—but not in the baryon-to-photon ratio—on length scales of order $1/(10H)$ (see section\ref{sec:lisa}). In this case, only modes with $\lambda_\nu k_{\rm phys} \gtrsim 1$, which are strongly damped, give rise to appreciable inhomogeneities in the baryon-to-photon ratio after the temperature fluctuations are smoothed out. This requires initial modes with physical wave vector $k_{\rm phys}$ satisfying
\beq
\frac{k_{\rm phys}}{H} \gtrsim \frac{1}{\lambda_\nu H} \sim \left(\frac{T}{T_\nu}\right)^3\,.
\eeq
Thus, for phase transitions confined to a purely dark sector, we expect BBN constraints to arise only if the transition occurs close to the MeV scale. By contrast, current data favors a transition near T $\sim 100 \, {\rm MeV}$, which would not lead to significant constraints from BBN.

Another possible explanation involves the collapse of a domain wall network~\cite{NANOGrav:2023hvm,Ferreira:2024eru}. After formation, the domain walls enter a scaling regime with only a few domains per Hubble volume. A small vacuum energy bias between domains eventually triggers the collapse of the network. To account for the PTA signal, the network must disappear around $T \sim 100~\mathrm{MeV}$ and contribute $\mathcal{O}(0.1)$ of the total energy density prior to collapse. For the collapse to complete efficiently, the energy bias between domains must also constitute a non-negligible fraction of the total energy density.

The released energy may be transferred either to the visible sector or to a purely dark sector. If it is transferred entirely into dark radiation, the scenario is in strong tension with recent constraints on $\Delta N_{\rm eff}$~\cite{Yeh:2022heq, ACT:2025tim}. By contrast, if the energy is deposited in the visible sector, the collapse generates inhomogeneities in the baryon-to-photon ratio. The spatial scales of these inhomogeneities are large enough to survive until BBN, though a precise determination of their magnitude requires dedicated numerical simulations, which we leave for future work.

On the other hand, if the energy remains confined to the dark sector, the collapse produces correlated fluctuations in temperature and baryon density, but not in the baryon-to-photon ratio. In this case, BBN constraints are expected only if the collapse occurs close to the MeV scale.

\section{Conclusions and outlook}\label{sec:conclusions}

In this paper, we investigated the constraints on baryon inhomogeneities imposed by Big Bang Nucleosynthesis (BBN). We demonstrated that the deuterium-to-hydrogen ratio, $D/H$, is particularly sensitive to spatial variations in the baryon-to-photon ratio, $\eta$. This sensitivity enables BBN to place novel bounds on the amplitude of baryon inhomogeneities at the time of nucleosynthesis.

By studying baryon number diffusion, we showed that inhomogeneities with length scales larger than the comoving horizon size of a radiation-dominated universe at $T \approx 3 \, \mathrm{TeV}$ survive until BBN. This scale is set by the proton diffusion length. We then considered two main physical regimes, defined by comparing the inhomogeneity length scale to the neutron diffusion length by the time of BBN:

\begin{itemize}
\item Both neutrons and protons remain inhomogeneous throughout BBN.
\item Neutrons homogenize before BBN, while protons do not.
\end{itemize}

For each regime, we quantified how spatial variations in $\eta$—or equivalently, variations in proton number density at fixed neutron density—affect the $D/H$ ratio. Using the \texttt{PRyMordial} code, we numerically evaluated the deuterium abundance as a function of the local inhomogeneity amplitude, then computed the spatially averaged abundance and compared it to observational constraints. Our main findings are: In the regime where both protons and neutrons remain inhomogeneous, current measurements of $D/H$ restrict the RMS amplitude of baryon inhomogeneities to be below $28\%$. If neutrons homogenize before BBN but protons do not, the $D/H$ abundance becomes even more sensitive to local inhomogeneities, leading to a tighter bound of $26\%$ on the RMS amplitude. The stronger constraint in the second regime is physically intuitive: when neutrons are uniformly distributed, inhomogeneities in the proton density at the time of deuterium freeze-out are larger. Since the proton density directly controls the efficiency of deuterium destruction, $D/H$ becomes more sensitive to initial spatial variations in $\eta$.

We also discussed a third regime—where neutron diffusion remains active during BBN. In this case, neutron diffusion smooths out secondary inhomogeneities generated during nucleosynthesis. As shown in appendix~\ref{app:diffusionduringBBN}, the outcome effectively reduces to that of the case without neutron diffusion, since diffusion during BBN reintroduces correlations between proton and neutron densities.

We have shown that our bounds place important constraints on scenarios that generate inhomogeneities in the baryon-to-photon ratio. These constraints apply both to models in which the baryon asymmetry is produced inhomogeneously and to those where inhomogeneities are imprinted onto an initially homogeneous baryon-to-photon ratio. In the context of baryogenesis, we explored the potential reach of our results to constrain electroweak baryogenesis and discussed two recently proposed scenarios that are in tension with our bounds. We also emphasized the complementarity between our bounds and gravitational wave observations: first-order phase transitions capable of producing a sizable gravitational wave signal also imprint inhomogeneities in the baryon-to-photon ratio. In particular, we showed that pre-BBN sources generating a large stochastic GW background in the frequency range $\text{pHz} - 0.1 \,\text{mHz}$ can be probed through BBN.

Next, we provided estimates for the reach of this probe in the context of the electroweak phase transition, electroweak baryogenesis, and scenarios that source gravitational waves. We also noted that models producing primordial black holes typically generate sizable inhomogeneities. A precise determination of the constraints in such scenarios, however, requires dedicated analyses, which we leave for future work.

Finally, we emphasize the central role of nuclear cross sections in setting the bounds presented here. While substantial improvements are anticipated both in the measurements of $D/H$ and in the CMB determination of $\eta$, these will translate into stronger constraints only if the precision of the relevant nuclear rates improves in parallel. The most critical reactions are those that deplete deuterium, in particular $D + D \to T + p, D + D \to {}^3\mathrm{He} + n$, and $D + p \to {}^3\mathrm{He} + \gamma$. The recent improvement in the precision of the $D + p \to {}^3\mathrm{He} + \gamma$ rate by the LUNA collaboration has already enhanced the accuracy of the theoretical prediction for $D/H$. Nevertheless, further refinement is required to match the expected advances in the precision of $D/H$ and $\eta$ measurements.

\acknowledgments
We would like thank Peizhi Du, Andrew Gomes, Oleksii Matsedonskyi, Riccardo Rattazzi, and Marc Riembau, for discussions. We thank Matt Reece for comments on an early draft of this paper. We would like to thank Miguel Escudero for discussions and a BBN course at the University of Geneva. We  also  thank Anne-Katherine Burns for guidance on modifying the PRyMordial code. HB is supported by the DOE Grant DE-SC0013607.
ME and SS are supported by
the Swiss National Science Foundation under contract 200020-213104. ME and SS acknowledge the hospitality of the CERN theory group.

\vspace{0.5 cm}

\newpage

\appendix

\section{Toy model for inhomogeneities}
\label{app:toymodelinhomogeneities}
In this appendix, we present a toy model for inhomogeneities and their subsequent evolution under diffusion. We show that the variance of the final fluctuations is determined by the interplay between the characteristic length scale of the initial density profile and the diffusion length.

Consider an initial density profile localized in small regions, separated by distances much larger than their individual sizes. This setup can be modeled by defining an initial density profile over a cubic region of volume $(N L)^3$, given by

\beq
\rho(\vec{r},t=0)= \frac{1}{N^3} \sum_{n_x,n_y,n_x =-N/2}^{N/2} \delta^3(\vec{r}-L(n_x \hat{x}+ n_y \hat{z} + n_z \hat{z}) )\,,
\eeq
where the sum runs over integer values of $n_x$, $n_y$, $n_z$, and $N$ is a large integer with the limit $N \to \infty$ taken at the end. The normalization is chosen such that the total charge over the entire space is unity. At later times, under diffusion, the density profile evolves into
\beq
\rho(\vec{r},t)= \frac{1}{(\sqrt{2 \pi} N d(t))^3} \sum_{\vec{n}} e^{-\frac{(\vec{r}-L \vec{n})}{2 d(t)^2}}\,.
\eeq
with $d(t)$ being the diffusion length at time $t$. 

\begin{figure}[t]
    \centering
    \includegraphics[width=0.6\linewidth]{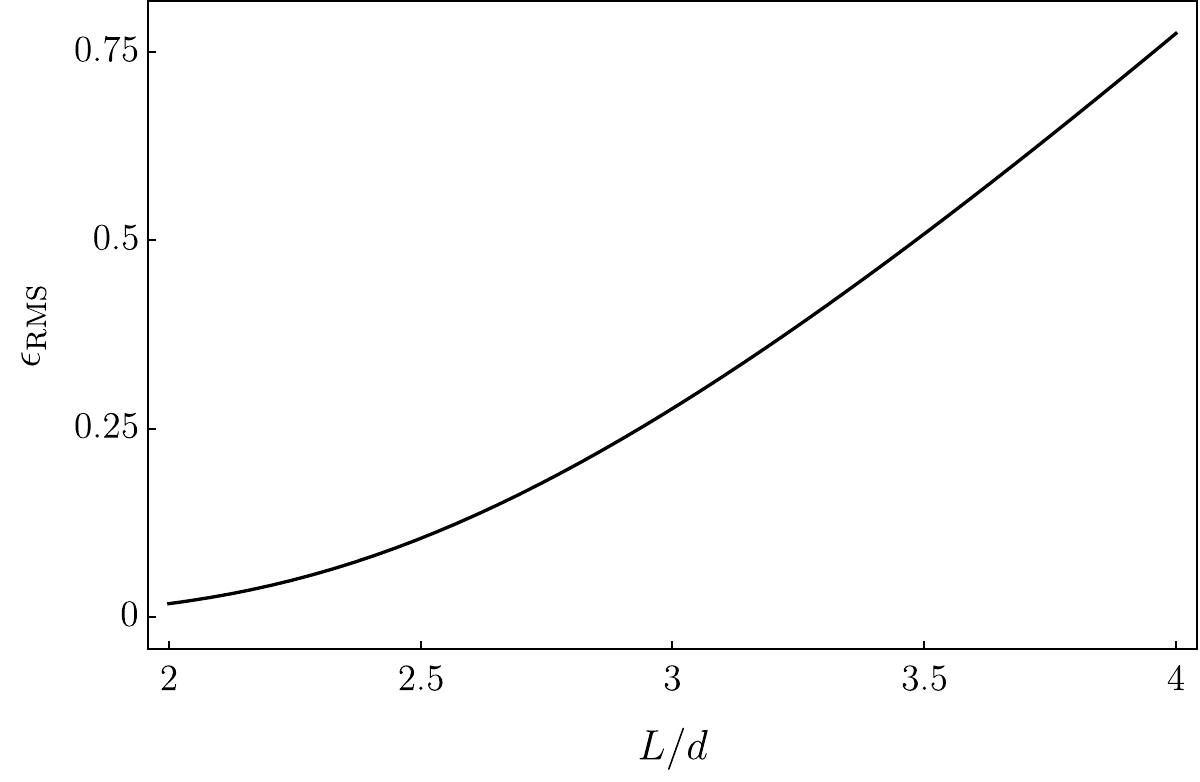}
    \caption{RMS fluctuation $\epsilon_\text{RMS}$ as a function of $L/d$.}
    \label{fig:epsrmsasfunctionofLoverd}
\end{figure}

We are interested in evaluating the RMS fluctuation, $\epsilon_\text{RMS}$, defined as
\beq
\epsilon_\text{RMS}^2 = \frac{\langle \rho^2 \rangle - \langle \rho \rangle^2}{\langle \rho \rangle^2}\,.
\eeq
To compute $\langle \rho^2 \rangle$, it is convenient to work in momentum space:
\beq
\int d^3x \, \rho^2(\vec{r},t) = \int \frac{d^3k}{(2\pi)^3}\, \big|\tilde{\rho}(\vec{k},t)\big|^2\,,
\eeq
with $\tilde{\rho}(\vec{k},t)$ the Fourier transform of $\rho(\vec{r},t)$. This takes the form
\beq
\tilde{\rho}(\vec{k},t) = \frac{1}{N^3} \sum_{\vec{n}} e^{i \vec{k}\cdot \vec{n}L} \, e^{-\tfrac{1}{2} d(t)^2 k^2}\,.
\eeq
From this expression, we obtain
\beq
\epsilon_\text{RMS}^2 = \left( \frac{L}{2\sqrt{\pi}\, d(t)} \sum_{n=-\infty}^{\infty} e^{-\tfrac{L^2}{4 d(t)^2} n^2} \right)^{3} - 1\,,
\eeq
after taking the limit $N \to \infty$, corresponding to infinite volume. The resulting dependence on $L/d$ is shown in figure~\ref{fig:epsrmsasfunctionofLoverd}. At late times, where $d$ becomes large, the distribution approaches homogeneity and $\epsilon_\text{RMS}$ vanishes.

In this example, we considered only the effect of diffusion, modeled with a Gaussian transfer function. In this case, the limit $\epsilon_\text{RMS}=0.3$ is reached at $L/d\simeq 3$, and $\epsilon_\text{RMS}$ then falls exponentially as $L/d$ further decreases.

The situation changes when the evolution of inhomogeneities in the proton number density is taken into account. As shown in section~\ref{sec:diffusionrefined}, the corresponding transfer function deviates significantly from a pure Gaussian, in particular exhibiting a slower falloff for momenta larger than $d_p^{-1}$. As a result, $\epsilon_\text{RMS}$ remains larger at small $L/d$.

\section{Diffusion during BBN}
\label{app:diffusionduringBBN} 

\begin{figure}[t]
    \centering
    \includegraphics[width=0.6\linewidth]{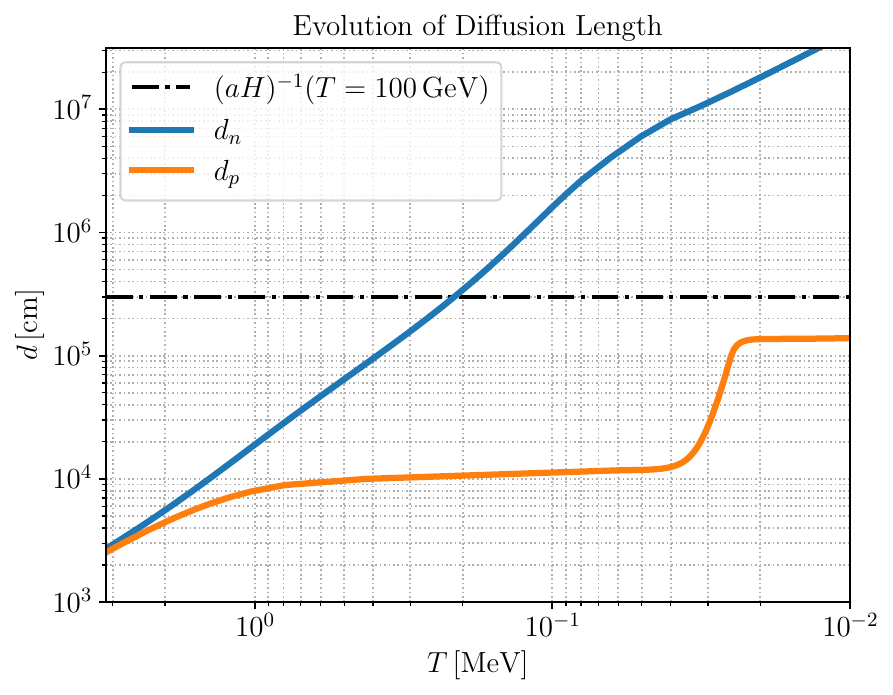}
    \caption{Proton and neutron comoving diffusion lengths up to the end of BBN.}
    \label{fig:lowTdiffusioncomb}
\end{figure}

In this section, we examine the role of diffusion during BBN. We begin with neutron diffusion and then turn to the case of proton diffusion. While protons do not diffuse appreciably between neutrino decoupling and the onset of BBN, they begin to do so once the temperature drops below $T \lesssim 30~\mathrm{keV}$. As we show, this has only a minor impact on the final light-element abundances. The diffusion lengths of protons and neutrons up to the end of BBN are displayed in figure~\ref{fig:lowTdiffusioncomb}.

\subsection{Neutron diffusion during BBN} \label{app:neutronDBNN}

We focus here on the regime in which neutrons diffuse and homogenize while protons remain inhomogeneous. In this situation, BBN proceeds more rapidly in regions of higher proton density. Deuterium forms more efficiently in these regions, leading to a faster depletion of neutrons, which then diffuse in order to re-establish homogeneity.

We now turn to the case in which neutron diffusion remains active during BBN. Our aim is to estimate the final proton abundance, starting from an initially inhomogeneous proton distribution together with a homogeneous neutron distribution. During BBN, nearly all neutrons are incorporated into ${}^4\mathrm{He}$, predominantly through two fusion chains:
\begin{align}
   & 3p + 3n \to 3D \to T + D + p \to {}^4\mathrm{He} + p + n\,,\\ 
   & 3p + 3n \to 3D \to D + {}^3\mathrm{He} + n \to D + T + p \to {}^4\mathrm{He} + p + n\,.
\end{align}
The initial production of deuterium proceeds slowly, but once a sufficient abundance is established, subsequent reactions occur rapidly. As a result, the abundances of deuterium, ${}^3He$, and tritium can be treated as being in quasi-equilibrium. The effective reaction governing the evolution of neutron and proton densities is then:
\begin{equation}
    3n + 3p \to {}^4\mathrm{He} + n + p\,.
\end{equation}
The evolution equations for the comoving neutron and proton densities are then:
\begin{align} \label{eq:neutrondiffduringBBN}
    \partial_t n_n(x,t) &= - \frac{2}{3} \gamma(t) n_p(x,t) n_n(x,t) + D \nabla^2 n_n(x,t)\,, \\
    \partial_t n_p(x,t) &= - \frac{2}{3} \gamma(t) n_p(x,t) n_n(x,t)\,,
\end{align}
where $\gamma(t)$ sets the deuterium production rate. The factor of $2/3$ accounts for the need to form three deuterium nuclei per ${}^4\mathrm{He}$ nucleus (along with a leftover neutron and proton).

Inhomogeneities in the proton density enhance reaction rates in overdense regions, which in turn drive additional neutron diffusion. To simplify the analysis, we consider the limit $D \gg \gamma\, n_p(x,t)\, L^2$, where $L$ denotes the characteristic length scale of the inhomogeneities. In this regime, the diffusion rate is much larger than the rate at which neutrons are consumed through deuterium formation, allowing us to treat neutron diffusion as effectively instantaneous:
\begin{equation}
    n_n(x,t) = n_n(t)\,.
\end{equation}
Let us consider at some initial time $t_i$ just before the onset of helium formation two homogeneous regions with initial proton densities $ n_p^+(t_i) = \bar{n}_p (1 + \epsilon)$ and $n_p^-(t_i) = \bar{n}_p (1 - \epsilon)$ where $\bar{n}_p$ is the average proton density. Their evolution is governed by
\begin{align}
    \partial_t n_p^+(t) &= -\frac{2}{3} \gamma(t) n_p^+(t) n_n(t)\,, \\
    \partial_t n_p^-(t) &= -\frac{2}{3} \gamma(t) n_p^-(t) n_n(t)\,.
\end{align}
These equations are linear in the proton number densities, and the consumption rate of protons in two regions are equal. Solving these equations gives
\begin{equation}
    n_p^+(t) - n_p^-(t) = \left(n_p^+(t_i) - n_p^-(t_i)\right) \exp\left[ - \int_{t_i}^t dt' \frac{2}{3} \gamma(t') n_n(t') \right]\,. \label{eq:protoninhom}
\end{equation}
To evaluate the exponential factor, we can examine the average proton density
\begin{equation}
    \partial_t \left( \frac{n_p^+(t) + n_p^-(t)}{2} \right) = -\frac{2}{3} \gamma(t) \left( \frac{n_p^+(t) + n_p^-(t)}{2} \right) n_n(t)\,. \label{eq:neutronstimeevol}
\end{equation}
Since protons are mainly depleted through fusion with neutrons, the total difference between proton and neutron numbers is a constant,
\begin{equation}
    \frac{n_p^+(t) + n_p^-(t)}{2} - n_n(t) = c = \text{const}\,.
\end{equation}
It is important to note that the constant $c$ is only determined by the space-averaged number densities and is therefore independent of the amount of inhomogeneity in the initial conditions. Now we can plug this back into equation~\eqref{eq:neutronstimeevol} to find
\begin{equation}
    \partial_t n_n(t) = - \frac{2}{3} \gamma(t) \left(c + n_n(t)\right) n_n(t)\,.
\end{equation}

This shows that evolution of the neutron number density in this regime is independent of whether the initial proton distribution is homogeneous or not. We can therefore find it by solving the equations for the homogeneous case. In that case, the proton density evolves as
\begin{equation}
    \partial_t n_p(t) = -\frac{2}{3} \gamma(t) n_p(t) n_n(t)\,,
\end{equation}
with solution
\begin{equation}\label{eq:protonsolution}
    n_p(t) = n_p(t_i) \exp\left[ - \int_{t_i}^t dt' \frac{2}{3} \gamma(t') n_n(t') \right]\,.
\end{equation}
The exponential can now be expressed in terms of the proton mass fraction $X_p$ as
\begin{equation}
    \exp\left[ - \int_{t_i}^t dt' \frac{2}{3} \gamma(t') n_n(t') \right] = \frac{X_p(T)}{X_p(T_i)}\,,
\end{equation}
where $T_i\approx 60 \, \rm{MeV}$ is the temperature at onset of BBN. If we consider a time $t_f$ at the end of helium formation we then have 
\begin{equation}
    \frac{n_p(t_f)}{n_p(t_i)} = \frac{1 - 2X_n(T_i)}{1 - X_n(T_i)}\,. 
\end{equation}
Plugging this into equation~\eqref{eq:protoninhom}, we find:
\begin{equation}
    n_p^+(t_f) - n_p^-(t_f) = \left(n_p^+(t_i) - n_p^-(t_i)\right) \frac{1 - 2X_n(T_i)}{1 - X_n(T_i)} = 2 \epsilon \, \bar{n}_p(t_i) \frac{1 - 2X_n(T_i)}{1 - X_n(T_i)}\,.
\end{equation}

Let's compare this with the regime where the length scale of inhomogeneities are larger than the neutron diffusion length so that we can consider the inhomogeneities to be correlated and captured by that of $\eta$. In this case the difference in the proton density between the two patches after ${}^4\mathrm{He}$ formation is
\begin{equation}
    n_p^+(t_f) - n_p^-(t_f) = \left(n_p^+(t_i) - n_n^+(t_i)\right) - \left(n_p^-(t_i) - n_n^-(t_i)\right) = 2 \epsilon \, \bar{n}_p(t_i)\frac{1 - 2X_n(T_i)}{1 - X_n(T_i)}\,.
\end{equation}
In the above, the first equality expresses the fact that all neutrons are consumed in the formation of ${}^4\mathrm{He}$ within each patch, while the second equality follows from $\bar{n}_p(t_i) - \bar{n}_n(t_i) = \bar{n}_p(t_i)\,(1 - 2X_n(T_i))/(1 - X_n(T_i))$. This shows that, in the limit of rapid neutron diffusion, the deuterium abundance remains just as sensitive to baryon inhomogeneities as in the case without diffusion.

\subsection{Proton diffusion during BBN}\label{app:protonDBBN}

Following neutrino decoupling, the proton diffusion coefficient remains small enough that proton diffusion is negligible. This situation persists until the electron density becomes exponentially suppressed, at which point the proton diffusion coefficient rises rapidly as the temperature decreases. In this regime, proton diffusion is ultimately limited by electron diffusion, since the two are coupled in order to maintain charge neutrality. We conclude this section with a discussion of the correspondingly small impact of proton diffusion on the final deuterium abundance.

\subsubsection{Proton diffusion length during BBN}

We begin by outlining how electron diffusion ultimately limits the diffusion of protons. This occurs through the build-up of an electric potential, which induces an electric current carried by protons that is comparable to the proton diffusion current. At high temperatures, however, the dense electron–positron plasma efficiently screens the Coulomb potential over short distances. As a result, the electric force acting on protons is negligible, and proton diffusion can be treated independently, as assumed in the main text.

In this subsection, we study the evolution of the proton and electron number densities on sufficiently short length and time scales that the Hubble expansion can be neglected. The coupled evolution equations governing these number densities are given by (see, e.g., the discussion of ambipolar diffusionin~\cite{Pitaevskii1981Physical}):
\begin{align}
    \frac{\partial \mathfrak{n}_p}{\partial t} &= D_p \left[ \nabla^2 \mathfrak{n}_p - \frac{e}{T} \vec{\nabla} \cdot (\vec{E} \, \mathfrak{n}_p) \right]\,, \label{eq:coupledelectronproton} \\
    \frac{\partial \delta \mathfrak{n}_e}{\partial t} &= D_e \left[ \nabla^2 \delta \mathfrak{n}_e + \frac{e}{T} \vec{\nabla} \cdot (\vec{E} \, (\mathfrak{n}_{e^-}+\mathfrak{n}_{e^+})) \right]\,.
    \label{eq:coupledelectronproton2}
\end{align}
In these equations, $\delta \mathfrak{n}_e = \mathfrak{n}_{e^-} - \mathfrak{n}_{e^+}$ denotes the net electron number density in proper volume units, $D_e$ is the electron diffusion coefficient, and $\vec{E}$ is the electric field generated by the charge distribution,
\begin{equation}
    \vec{E} = - \vec{\nabla} \phi \qquad \text{, with} \qquad \nabla^2 \phi = - e \left(\mathfrak{n}_p - \delta \mathfrak{n}_e \right)\,,
\end{equation}
where $\Phi$ is the electric potential. 

Typically, the electron diffusion coefficient is much smaller than the proton diffusion coefficient and is dominated by electron–photon scattering. In eqs.~\eqref{eq:coupledelectronproton} and~\eqref{eq:coupledelectronproton2}, the first term represents the diffusion of protons and electrons, respectively, while the second term encodes their response to the electric potential. These equations simplify considerably for approximately homogeneous distributions, since $\vec{\nabla} \cdot \big(\vec{E}\,\mathfrak{n}_p\big) \simeq \bar{\mathfrak{n}}_p \, \vec{\nabla} \cdot \vec{E}$ and similarly for electrons, with $\bar{\mathfrak{n}}$ denoting the average density. Under this approximation, and working in momentum space, we obtain:
\begin{align}
    \frac{\partial \tilde{\mathfrak{n}}_p}{\partial t} &= D_p \left[ -k^2 \tilde{\mathfrak{n}}_p - k_{D,p}^2 (\tilde{\mathfrak{n}}_p - \delta \tilde{\mathfrak{n}}_e) \right]\,, \label{eq:coupledelectronprotonmom} \\
    \frac{\partial \delta \tilde{\mathfrak{n}}_e}{\partial t} &= D_e \left[ -k^2 \delta \tilde{\mathfrak{n}}_e + k_{D,e}^2 (\tilde{\mathfrak{n}}_p + \delta \tilde{\mathfrak{n}}_e) \right]\,. 
    \label{eq:coupledelectronprotonmom2}
\end{align}
We have defined $k^2_{D,p} \equiv \frac{ e^2}{T} \bar{\mathfrak{n}}_p$ and $k^2_{D,e} \equiv \frac{ e^2}{T} \left(\bar{\mathfrak{n}}_{e^-} + \bar{\mathfrak{n}}_{e^+} \right)$ as the square of the Debye wave-vector for protons and electrons respectively.

For the length scales of interest, we always have $k\ll k_{D}$ for both protons and electrons. 
Subtracting the two equations above, we find the following equation governing the evolution of the electric charge density 
\begin{align}\label{eq:coupledsubtracted}
    \frac{\partial \left(\tilde{\mathfrak{n}}_p-\delta \tilde{\mathfrak{n}}_e\right)}{\partial t} &= -\left[D_p k_{D,p}^2 + D_e k^2_{D,e} \right] \left( \tilde{\mathfrak{n}}_p - \delta \tilde{\mathfrak{n}}_e \right) - \left( D_p \tilde{\mathfrak{n}}_p -D_e \delta \tilde{\mathfrak{n}}_e\right) k^2\,.
\end{align}
A fluctuation in charge density would be exponential damped over a short time scale $\tau$ of order
\beq \tau\sim \left[D_p k_{D,p}^2 + D_e k^2_{D,e} \right]^{-1}\,.
\eeq
This behavior continues until the second term in equation~\eqref{eq:coupledsubtracted} becomes non-negligible. At this point there is a residual charge density of order
\begin{equation}\label{eq:chargedensity}
     \left( \tilde{\mathfrak{n}}_p - \delta \tilde{\mathfrak{n}}_e \right) \sim \frac{ D_p k^2 }{D_p k_{D,p}^2 + D_e k^2_{D,e}} \tilde{\mathfrak{n}}_p\,,
\end{equation}
where we used $D_p\gg D_e$ and $\tilde{\mathfrak{n}}_p \approx \delta \tilde{\mathfrak{n}}_e$. This equation shows that the residual electric charge density is much smaller than the separate electron and proton densities. 

We now distinguish between two regimes. At high temperatures, the electron density is unsuppressed, leading to $D_e k^2_{D,e} \gg D_p k^2_{D,p}$, while at low temperatures the opposite holds, $D_e k^2_{D,e} \ll D_p k^2_{D,p}$. In the high-temperature regime, substituting equation~\eqref{eq:chargedensity} into the proton diffusion equation~\eqref{eq:coupledelectronprotonmom}, we obtain
\beq
 \frac{\partial \tilde{\mathfrak{n}}_p}{\partial t} = - D_p k^2 \tilde{\mathfrak{n}}_p \left[ 1+ \mathcal{O} \left( \frac{ D_p k_{D,p}^2 }{ D_e k^2_{D,e}}\right) \right].
\eeq
Thus, at high temperatures the electric force on protons can be neglected, and proton diffusion proceeds independently of electron diffusion. This situation changes once the electron density becomes sufficiently small that $D_e k^2_{D,e} \sim D_p k^2_{D,p}$.

In the low-temperature regime, where $D_e k^2_{D,e} \ll D_p k^2_{D,p}$, the coupled equations once again simplify. In this case, both terms on the right-hand side of the proton diffusion equation~\eqref{eq:coupledelectronprotonmom} are of comparable size. Since the proton diffusion coefficient is much larger than that of the electrons, protons rapidly approach equilibrium, where the two contributions cancel. In other words, we find
\begin{equation}
     \left( \tilde{\mathfrak{n}}_p - \delta \tilde{\mathfrak{n}}_e \right) \simeq \frac{k^2 }{k_{D,p}^2} \tilde{\mathfrak{n}}_p\,.
\end{equation}
Plugging this into the electron diffusion equation~\eqref{eq:coupledelectronprotonmom2}, we get
\beq
 \frac{\partial \delta \tilde{\mathfrak{n}}_e}{\partial t} = - D_e k^2 \delta \tilde{\mathfrak{n}}_e \left[ 1+ \frac{k_{D,e}^2}{k_{D,p}^2}\right]\,.
\eeq
Meaning, electrons diffuse with an effective diffusion coefficient of 
\begin{equation}\label{eq:Deff}
    D^\text{eff} = D_e \left[ 1+ \frac{k_{D,e}^2}{k_{D,p}^2}\right]\,.
\end{equation}

Since $|\tilde{\mathfrak{n}}_p - \tilde{\mathfrak{n}}_e| \ll \tilde{\mathfrak{n}}_p$, the same equations apply to the proton density. Consequently, charge neutrality ensures that protons also diffuse with the effective diffusion coefficient given in~\eqref{eq:Deff}. In the very low-temperature regime, this coefficient saturates at $D^\text{eff} = 2D_e$. We note that this result is consistent with the findings of~\cite{Pitaevskii1981Physical}, where the effective diffusion coefficient becomes twice the smaller of the two diffusion coefficients.

At the temperatures of interest, the electron diffusion is dominated by scattering of photons\footnote{At low temperatures, electron densities are exponentially suppressed, which might lead one to expect that proton-photon scattering is the dominant contribution to the proton diffusion coefficient. However, this is not the case; proton-photon scattering is suppressed relative to electron-photon scattering by a factor of $\left(m_e/m_p\right)^2$ for all temperatures of interest and is therefore negligible.}. The diffusion coefficient is given by
\begin{equation}
    D_e^{-1} = \frac{4 \pi^2}{45} \sigma_{Th} T^3,
\end{equation}
where $\sigma_{Th}= \frac{8\pi}{3}\frac{\alpha^2}{m_e^2}$ is the Thomson scattering cross section. The evolution of both the proton and neutron diffusion lengths after the on-set of BBN are shown in figure~\ref{fig:lowTdiffusion}.

\subsubsection{Implications of proton diffusion during BBN for deuterium abundance}

\begin{figure}
\centering
\begin{minipage}[c]{0.49\textwidth}
\centering
\includegraphics[width=\columnwidth]{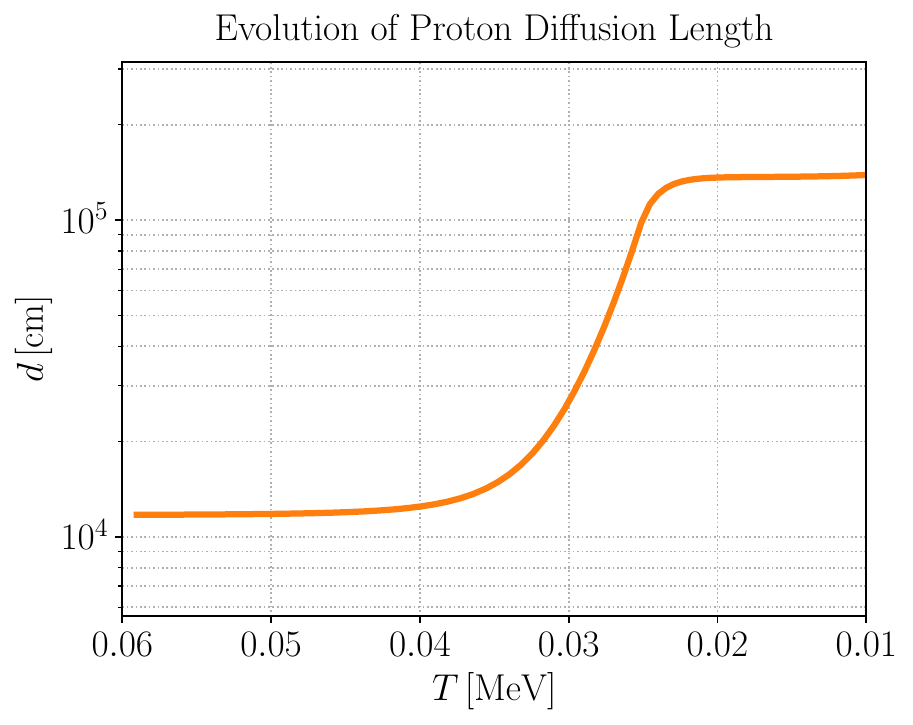}
\end{minipage}
\hfill
\begin{minipage}[c]{0.49\textwidth}
\centering
\includegraphics[width=\columnwidth]{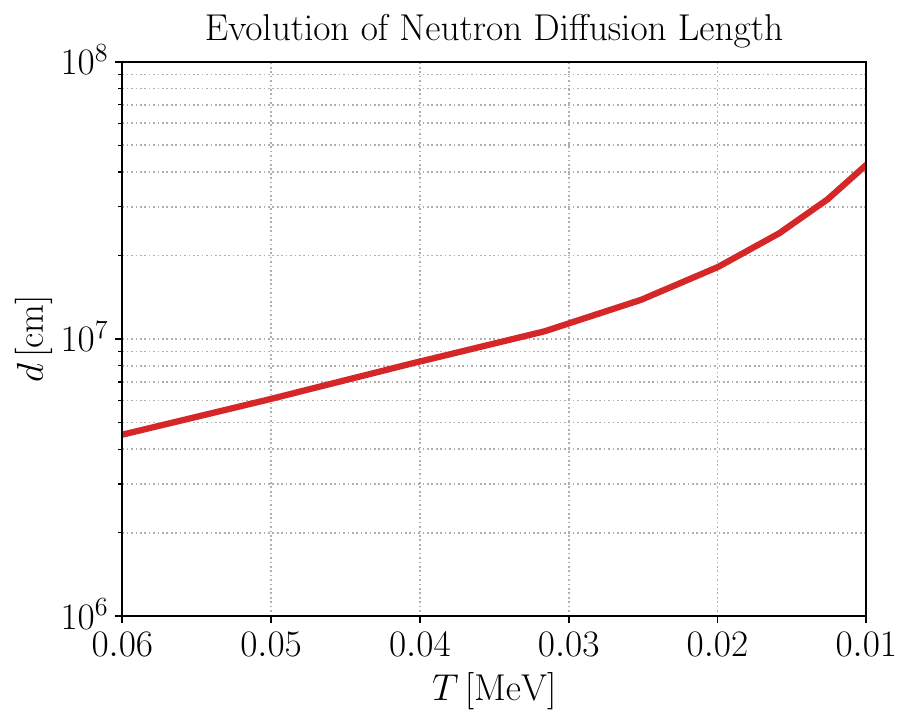}
\end{minipage}
\caption{Proton diffusion \textbf{(Left)} and neutron diffusion \textbf{(Right)} during BBN.} \label{fig:lowTdiffusion}
\end{figure}

As illustrated in figure~\ref{fig:lowTdiffusion}, proton diffusion during BBN is negligible for $T \gtrsim 35\,\mathrm{keV}$. From that point until the end of BBN at around $20\,\mathrm{keV}$, the proton diffusion length increases by roughly a factor of six. We now estimate the impact of this diffusion on the final deuterium abundance in the regime where the inhomogeneity length scale lies within this growth range—neither far above the proton diffusion length at $20\,\mathrm{keV}$ nor far below that at $35\,\mathrm{keV}$.

To place an upper bound on the impact of proton diffusion during BBN on inhomogeneity length scales within the growth region, we assume that protons and deuterium remain inhomogeneous down to a temperature $T_\text{diff}$ defined by
\begin{equation}
k^{-1} = d_p(T_\text{diff}) ,
\end{equation}
with $k^{-1}$ the characteristic comoving length scale of the inhomogeneities. At this temperature we approximate both species as instantaneously homogenized. Following this homogenization, the proton abundance coincides with that of standard homogeneous BBN, while the deuterium abundance is enhanced by $\mathcal{O}(\langle \epsilon^2 \rangle)$ relative to the homogeneous case, as discussed in the main text.

As discussed below equation~\eqref{eq:heab}, the final deuterium abundance is determined by the freeze-out of the deuterium-destroying reactions. For $T < 35\,\mathrm{keV}$, the deuterium abundance changes by only about 35\%. Treating this variation as a small parameter, one would therefore expect corrections to the final deuterium abundance—relative to the case with negligible proton diffusion—of order $0.35\,\epsilon^2$, which could be non-negligible. As we will show, however, the actual effect is somewhat smaller.

\begin{figure}
\centering
\begin{minipage}[c]{0.49\textwidth}
\centering
\includegraphics[width=\columnwidth]{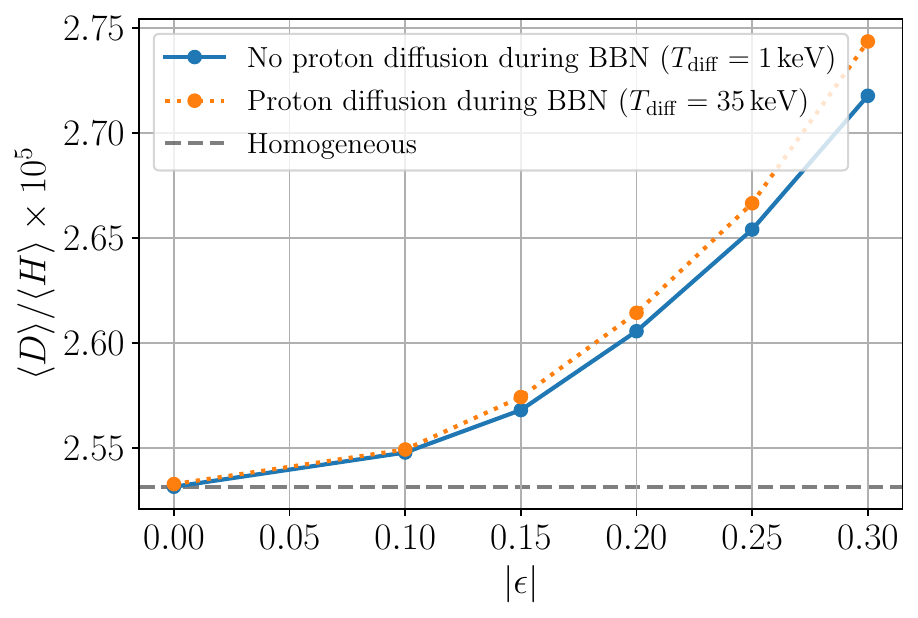}
\end{minipage}
\hfill
\begin{minipage}[c]{0.49\textwidth}
\centering
\includegraphics[width=\columnwidth]{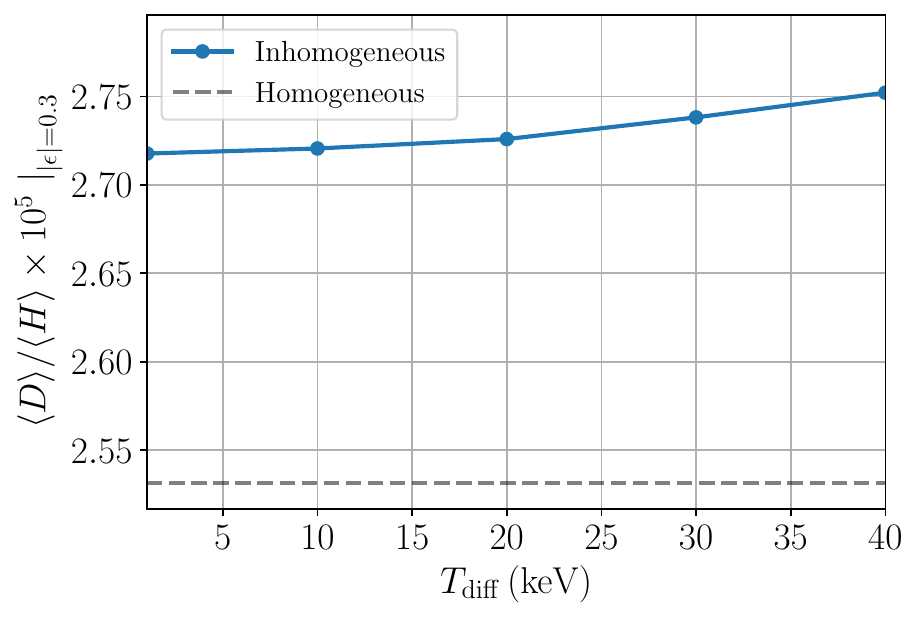}
\end{minipage}
\caption{Dependence of $\langle D\rangle/\langle H \rangle$ on the inhomogeneity magnitude $|\epsilon|$.
\textbf{(Left)} Results assuming instantaneous diffusion at $35\,\mathrm{keV}$ and $1\,\mathrm{keV}$. The dashed horizontal line corresponds to the homogeneous value.
\textbf{(Right)} For fixed $|\epsilon|=0.3$, the $\langle D\rangle/\langle H \rangle$ ratio shows varying sensitivity to the temperature at which diffusion is assumed to occur, with the strongest effect in the range $20\,\mathrm{keV}$–$40\,\mathrm{keV}$.} \label{fig:ep_dep}
\end{figure}

To quantify the impact of proton diffusion during BBN, we ran the \texttt{PRyMordial} code down to a chosen temperature $T_{\mathrm{diff}} < 40\,\mathrm{keV}$. At this point we assume instantaneous diffusion of neutrons, protons, and the light elements, and average their abundances across patches of positive and negative inhomogeneity. We then extract from the code the present-day value of the averaged deuterium-to-averaged hydrogen ratio. The corresponding results are shown in figure~\ref{fig:ep_dep}.

We conclude that if the inhomogeneities lie on sufficiently large length scales such that proton diffusion remains negligible until $20~\mathrm{keV}$, the constraints presented in the main text are essentially unchanged. On smaller length scales, however, the results can be modified by up to $15\%$. It should be emphasized that in this setup neutron diffusion between $100\,\mathrm{keV}$ and $T_{\mathrm{diff}}$ is neglected, so the calculation should be regarded only as an estimate of the effect of proton diffusion. A full quantification of this modification as a function of inhomogeneity length scale requires a combined treatment of BBN and diffusion, which we leave for future work.

\section{Evolution of sub-horizon temperature fluctuations}
\label{app:Tfluctuations}

In this appendix, we study the evolution of sub-horizon temperature fluctuations. We focus on both adiabatic modes, in which the baryon-to-photon ratio remains homogeneous, and baryon isocurvature modes, in which the baryon-to-photon ratio fluctuates. We show that these modes propagate in two distinct regimes: an overdamped regime, where strong damping of the temperature fluctuations leads to changes in the baryon-to-photon ratio, and an oscillatory regime, where the modes, though still damped and producing heat, leave the baryon-to-photon ratio unchanged.

Let us consider the evolution of temperature fluctuations. In general, such fluctuations propagate as sound waves and are damped both by viscosity and by heat diffusion. In the regime of interest, however, the effect of viscosity—set by the mean free path of particles strongly coupled to the fluid (e.g.~photons)—can be neglected compared to heat diffusion, which is governed by the much longer mean free path of neutrinos. More explicitly, let us consider a fluid with small perturbations in velocity, pressure, and temperature. The velocity responds to forces arising from the pressure gradient, leading to
\begin{equation}
  \rho \partial_t \vec{u} = - \vec{\nabla} p\,,
\end{equation}
and the pressure varies both in response to the compression or expansion of the fluid and due to heat diffusion,
\begin{equation}
    \partial_t p = - f(T) \vec{\nabla}\cdot \vec{u} + D(T) \nabla^2 p\,.
\end{equation}
Here $f(T)$ depends on the equation of state of the fluid and $D(T)$ is the heat diffusion constant.

\begin{figure}[t]
    \centering
\includegraphics[width=0.99\linewidth]{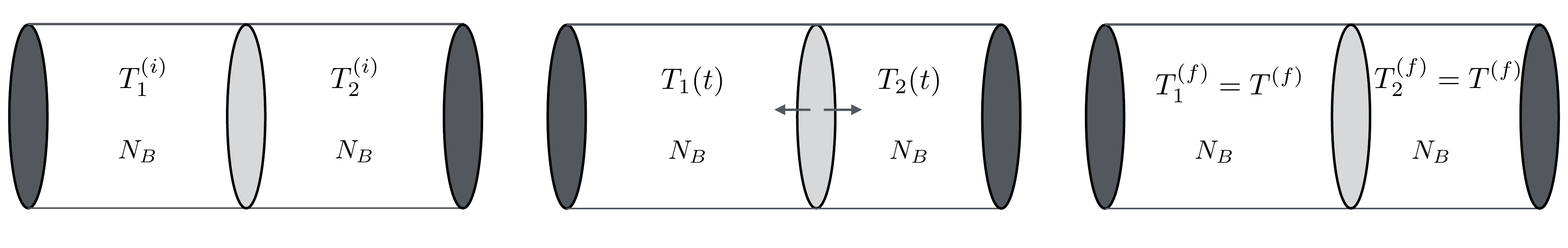}
    \caption{Time evolution of the fluid in the cylinder in the regime of slow heat diffusion (from left to right). The piston between the two fluids oscillates many times around its equilibrium value, before the oscillation is damped. As explained in the text, this does not lead to a change in the baryon-to-photon ratio.}
    \label{fig:slowheatdiff}
\end{figure}

Working in the Fourier space and assuming that the rate of change of $f(T)$ and $D(T)$ are both small compared to the frequency of the modes, gives the following dispersion relation
\begin{equation}
    \omega^2 = c_s^2 k^2 - i D(T) \omega k^2\,,
\end{equation}
where $c_s$ is the speed of sound. Thus, we find
\begin{equation} \label{eq:dispersion_fluid}
    \omega = \frac{i D(T) k^2}{2} \pm \sqrt{- \frac{D(T)^2 k^4}{4} + c_s^2 k^2}\,.
\end{equation}
This dispersion relation has two distinct regimes: (i) if $D(T)k \gg c_s$, diffusion dominates and the system is strongly damped; (ii) if $D(T)k \ll c_s$, the system undergoes many oscillations before significant damping occurs. To see the implication of these two regimes for the baryon-to-photon ratio, it is instructive to first study a toy example which captures the physics of this system.

\subsection*{A toy example}
Consider a cylinder of length $L$ that is divided into two regions by a frictionless piston. Each side of the cylinder contains the same number of baryons, denoted by $N_B$, as well as radiation with initial temperatures $T_1^{(i)}$ and $T_2^{(i)}$. Thus, the two sides begin with equal baryon number densities but different baryon-to-photon ratios. We further assume that the temperatures are sufficiently high that the radiation pressure dominates over the baryon pressure. The piston is impermeable to baryons but allows heat transfer between the two regions. For small temperature differences, the dispersion relation governing the motion of the piston takes the same form as equation~\eqref{eq:dispersion_fluid}. 

In the limit where heat transfer is very slow, the piston oscillates around the point at which the pressures—and hence the temperatures—on the two sides are equal. These oscillations are gradually damped by the small amount of heat exchange, and the piston ultimately settles at the equilibrium position. To determine this position, note that over a single oscillation period the heat transfer is negligible. On such time scales, for each side $j$, the relation $T_j^3 V_j = \mathrm{const.}$ holds, where $V_j$ is the corresponding volume. The piston therefore relaxes to a final position $x$, measured from the center of the cylinder, given by
\beq
x= \frac{L}{2} \frac{\left(\frac{T_1^{(i)}}{T_2^{(i)}}\right)^{1/3}-1}{\left(\frac{T_1^{(i)}}{T_2^{(i)}}\right)^{1/3}+1}\,.
\label{eq:xequilibrium}
\eeq
At the end of the process, the two sides settle with different baryon densities. However, the initial and final baryon-to-photon ratios on each side remain unchanged. This is because the net heat transfer to each side averages to zero over each oscillation cycle: during the half-cycle when $T_1 > T_2$, heat flows into the second region of the cylinder, while during the half-cycle when $T_2 > T_1$, the flow reverses. This situation is illustrated in figure~\ref{fig:slowheatdiff}, where, for concreteness, we take $T_1^{(i)} > T_2^{(i)}$. 

\begin{figure}[t]
    \centering
    \includegraphics[width=0.7\linewidth]{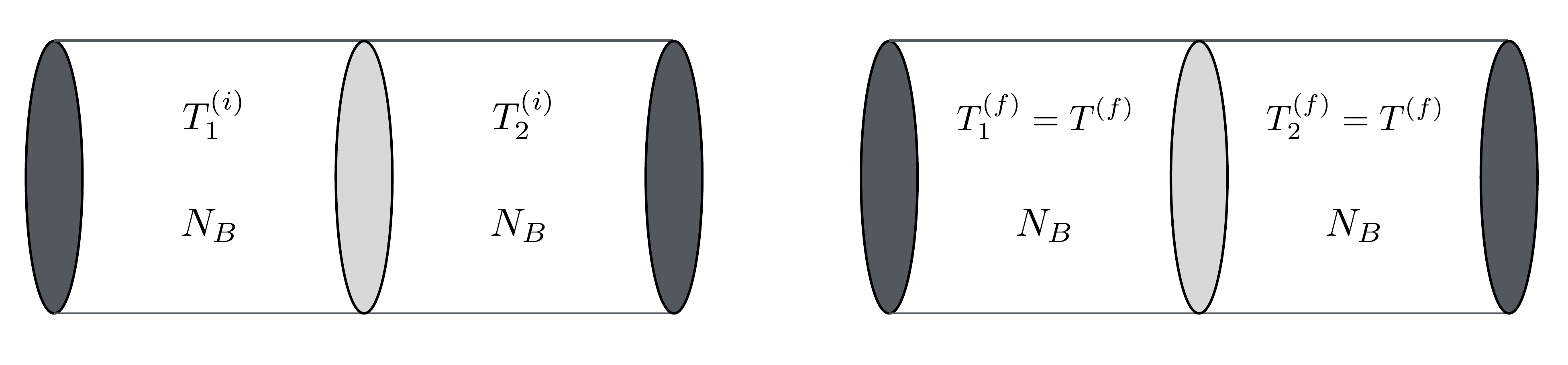}
    \caption{Time evolution of the fluid in the cylinder in the limit of infinite heat diffusion (from left to right). The heat is transferred before the piston starts to move, leading to a change in the baryon-to-photon ratio at both sides of the piston.}
    \label{fig:fastheatdiff}
\end{figure}

Now consider the opposite limit in which heat transfer is instantaneous. In this case, the two sides equilibrate without any motion of the piston, resulting in equal final temperatures and $x=0$. In this limit, the final baryon densities are the same, while the baryon-to-photon ratios differ. This situation is illustrated in figure~\ref{fig:fastheatdiff}.

If we move away from this limit and allow for finite but still rapid heat diffusion—such that the piston dynamics remain overdamped—the piston relaxes before reaching the equilibrium position given in equation~\eqref{eq:xequilibrium}. In this regime, heat flows only from the initially hotter side to the cooler side, and both the final baryon densities and the final baryon-to-photon ratios differ.

\subsection*{Evolution of temperature fluctuations}

The toy model above provides a good analogy for the evolution of fluctuations in the early universe. As discussed in the main text, baryon diffusion is slow, and the mean free paths of particles in the plasma (with the exception of neutrinos) are much smaller than the length scales of interest. Both of these features are effectively modeled by the impermeable piston in the example. Heat transfer prior to neutrino decoupling is dominated by neutrinos, whose mean free path can be macroscopic; we denote it by $\lambda_\nu$. From equation~\eqref{eq:dispersion_fluid}, we then estimate $D \sim \lambda_\nu$, while during radiation domination the sound speed satisfies $c_s^2 = 1/3$. Consequently, the condition for a mode with physical wavenumber $k_{\rm phys}$ to lie in the overdamped regime, $D > 2 c_s / k_{\rm phys}$, becomes
\beq
\lambda_\nu k_{\rm phys} \gtrsim 1\,.
\eeq
Below the electroweak scale $\lambda_\nu$ is be estimated as
\beq
\lambda_\nu \sim \left( G_F^2 T^5 \right)^{-1}\,,
\eeq
which is much smaller than the Hubble length scale at high $T$, but grows to be comparable to the Hubble length at the time of neutrino decoupling, $T_\nu \sim \mathrm{MeV}$.

We now turn to two illustrative examples:
\begin{enumerate}
\item Modes that are either generated at a scale not far above the horizon size or that become sub-horizon at some $T \gg T_\nu$.
\item Modes that are initially generated with $\lambda_\nu k_{\rm phys} \gtrsim 1$.
\end{enumerate}
In case (1), the modes start oscillating with negligible damping. Their amplitudes gradually decrease until the condition $D k_{\rm phys}^2 \sim H$ is met, at which point the amplitudes are significantly suppressed. Since at this stage the modes are well inside the horizon, $k_{\rm phys} \gg H$, it follows that $D k_{\rm phys} \ll 1$, meaning the modes remain underdamped. Thus, in this case, the baryon-to-photon ratio remains unchanged even after the modes have disappeared.

In case (2), however, the modes begin already in the overdamped regime and decay without oscillating. As a result, the baryon density is only slightly modified, while the final baryon-to-photon ratio does change. 

Finally, let us comment on some specific scenarios discussed in the main text. Scenarios in which inhomogeneities are produced at the electroweak scale—such as electroweak baryogenesis or a strong first-order electroweak phase transition—fall into case (1). In both situations, adiabatic as well as isocurvature modes are generated. Temperature inhomogeneities are damped in such a way that the baryon-to-photon ratio remains inhomogeneous. The situation differs for scenarios occurring near the MeV scale, such as phase transitions proposed to explain the PTA signal. If such a transition takes place solely within a dark sector, it produces inhomogeneities in temperature and baryon density, while the baryon-to-photon ratio remains homogeneous. However, if the transition is coupled to the visible sector, inhomogeneities in the baryon-to-photon ratio are also generated. For such late phase transitions, depending on the mode, the dynamics may correspond to case (2). In this regime the modes are strongly damped, and even the adiabatic fluctuations result in baryon-to-photon inhomogeneities, i.e. isocurvature modes.

\newpage
\bibliographystyle{JHEP}
\bibliography{biblio.bib}

\providecommand{\href}[2]{#2}\begingroup\raggedright\begin{thebibliography}{10}

\bibitem{Planck:2015fie}
{\scshape Planck} collaboration, \emph{{Planck 2015 results. XIII. Cosmological
  parameters}},
  \href{https://doi.org/10.1051/0004-6361/201525830}{\emph{Astron. Astrophys.}
  {\bfseries 594} (2016) A13}
  [\href{https://arxiv.org/abs/1502.01589}{{\ttfamily 1502.01589}}].

\bibitem{Planck:2018vyg}
{\scshape Planck} collaboration, \emph{{Planck 2018 results. VI. Cosmological
  parameters}},
  \href{https://doi.org/10.1051/0004-6361/201833910}{\emph{Astron. Astrophys.}
  {\bfseries 641} (2020) A6}
  [\href{https://arxiv.org/abs/1807.06209}{{\ttfamily 1807.06209}}].

\bibitem{Yeh:2022heq}
T.-H.~Yeh, J.~Shelton, K.A.~Olive and B.D.~Fields, \emph{{Probing physics
  beyond the standard model: limits from BBN and the CMB independently and
  combined}}, \href{https://doi.org/10.1088/1475-7516/2022/10/046}{\emph{JCAP}
  {\bfseries 10} (2022) 046}
  [\href{https://arxiv.org/abs/2207.13133}{{\ttfamily 2207.13133}}].

\bibitem{Mossa:2020gjc}
V.~Mossa et~al., \emph{{The baryon density of the Universe from an improved
  rate of deuterium burning}},
  \href{https://doi.org/10.1038/s41586-020-2878-4}{\emph{Nature} {\bfseries
  587} (2020) 210}.

\bibitem{Planck:2018jri}
{\scshape Planck} collaboration, \emph{{Planck 2018 results. X. Constraints on
  inflation}}, \href{https://doi.org/10.1051/0004-6361/201833887}{\emph{Astron.
  Astrophys.} {\bfseries 641} (2020) A10}
  [\href{https://arxiv.org/abs/1807.06211}{{\ttfamily 1807.06211}}].

\bibitem{Inomata:2016uip}
K.~Inomata, M.~Kawasaki and Y.~Tada, \emph{{Revisiting constraints on small
  scale perturbations from big-bang nucleosynthesis}},
  \href{https://doi.org/10.1103/PhysRevD.94.043527}{\emph{Phys. Rev. D}
  {\bfseries 94} (2016) 043527}
  [\href{https://arxiv.org/abs/1605.04646}{{\ttfamily 1605.04646}}].

\bibitem{Jeong:2014gna}
D.~Jeong, J.~Pradler, J.~Chluba and M.~Kamionkowski, \emph{{Silk damping at a
  redshift of a billion: a new limit on small-scale adiabatic perturbations}},
  \href{https://doi.org/10.1103/PhysRevLett.113.061301}{\emph{Phys. Rev. Lett.}
  {\bfseries 113} (2014) 061301}
  [\href{https://arxiv.org/abs/1403.3697}{{\ttfamily 1403.3697}}].

\bibitem{Buckley:2025zgh}
M.R.~Buckley, P.~Du, N.~Fernandez and M.J.~Weikert, \emph{{General Constraints
  on Isocurvature from the CMB and Ly-$\alpha$ Forest}},
  \href{https://arxiv.org/abs/2502.20434}{{\ttfamily 2502.20434}}.

\bibitem{Elor:2024cea}
G.~Elor, R.~Houtz, S.~Ipek and M.~Ulloa, \emph{{The Standard Model CP Violation
  is Enough}},  \href{https://arxiv.org/abs/2408.12647}{{\ttfamily
  2408.12647}}.

\bibitem{Azzola:2024pzq}
J.~Azzola, O.~Matsedonskyi and A.~Weiler, \emph{{Minimal Electroweak
  Baryogenesis via Domain Walls}},
  \href{https://arxiv.org/abs/2412.10495}{{\ttfamily 2412.10495}}.

\bibitem{Applegate:1985qt}
J.H.~Applegate and C.J.~Hogan, \emph{{Relics of Cosmic Quark Condensation}},
  \href{https://doi.org/10.1103/PhysRevD.31.3037}{\emph{Phys. Rev. D}
  {\bfseries 31} (1985) 3037}.

\bibitem{Applegate:1987hm}
J.H.~Applegate, C.J.~Hogan and R.J.~Scherrer, \emph{{Cosmological Baryon
  Diffusion and Nucleosynthesis}},
  \href{https://doi.org/10.1103/PhysRevD.35.1151}{\emph{Phys. Rev. D}
  {\bfseries 35} (1987) 1151}.

\bibitem{1990ApJ...358...36M}
G.J.~{Mathews}, B.S.~{Meyer}, C.R.~{Alcock} and G.M.~{Fuller}, \emph{{Coupled
  Baryon Diffusion and Nucleosynthesis in the Early Universe}},
  \href{https://doi.org/10.1086/168961}{\emph{APJ} {\bfseries 358} (1990) 36}.

\bibitem{Kurki-Suonio:1992knt}
H.~Kurki-Suonio, M.B.~Aufderheide, F.~Graziani, G.J.~Mathews, B.~Banerjee,
  S.M.~Chitre et~al., \emph{{Diffusion coefficients and inhomogeneous big bang
  nucleosynthesis}},
  \href{https://doi.org/10.1016/0370-2693(92)91207-P}{\emph{Phys. Lett. B}
  {\bfseries 289} (1992) 211}.

\bibitem{Suh:1998nt}
I.-S.~Suh and G.J.~Mathews, \emph{{Finite temperature effects on cosmological
  baryon diffusion and inhomogeneous big bang nucleosynthesis}},
  \href{https://doi.org/10.1103/PhysRevD.58.123002}{\emph{Phys. Rev. D}
  {\bfseries 58} (1998) 123002}
  [\href{https://arxiv.org/abs/astro-ph/9805179}{{\ttfamily
  astro-ph/9805179}}].

\bibitem{Jedamzik:1993tcf}
K.~Jedamzik and G.M.~Fuller, \emph{{The Evolution of nonlinear subhorizon scale
  entropy fluctuations in the early universe}},
  \href{https://doi.org/10.1086/173788}{\emph{Astrophys. J.} {\bfseries 423}
  (1994) 33} [\href{https://arxiv.org/abs/astro-ph/9312063}{{\ttfamily
  astro-ph/9312063}}].

\bibitem{Jedamzik:1993dc}
K.~Jedamzik, G.M.~Fuller and G.J.~Mathews, \emph{{Inhomogeneous primordial
  nucleosynthesis: Coupled nuclear reactions and hydrodynamic dissipation
  processes}}, \href{https://doi.org/10.1086/173789}{\emph{Astrophys. J.}
  {\bfseries 423} (1994) 50}
  [\href{https://arxiv.org/abs/astro-ph/9312065}{{\ttfamily
  astro-ph/9312065}}].

\bibitem{Heckler:1993nc}
A.~Heckler and C.J.~Hogan, \emph{{Neutrino heat conduction and inhomogeneities
  in the early universe}},
  \href{https://doi.org/10.1103/PhysRevD.47.4256}{\emph{Phys. Rev. D}
  {\bfseries 47} (1993) 4256}.

\bibitem{Scherrer:2021tbo}
R.J.~Scherrer, \emph{{Does inhomogeneous big bang nucleosynthesis produce an
  inhomogeneous element distribution today?}},
  \href{https://doi.org/10.1103/PhysRevD.103.123548}{\emph{Phys. Rev. D}
  {\bfseries 103} (2021) 123548}
  [\href{https://arxiv.org/abs/2103.01832}{{\ttfamily 2103.01832}}].

\bibitem{Jedamzik:2001qc}
K.~Jedamzik and J.B.~Rehm, \emph{{Inhomogeneous big bang nucleosynthesis: Upper
  limit on Omega(b) and production of lithium, beryllium, and boron}},
  \href{https://doi.org/10.1103/PhysRevD.64.023510}{\emph{Phys. Rev. D}
  {\bfseries 64} (2001) 023510}
  [\href{https://arxiv.org/abs/astro-ph/0101292}{{\ttfamily
  astro-ph/0101292}}].

\bibitem{Kainulainen:1998vh}
K.~Kainulainen, H.~Kurki-Suonio and E.~Sihvola, \emph{{Inhomogeneous big bang
  nucleosynthesis in light of recent observations}},
  \href{https://doi.org/10.1103/PhysRevD.59.083505}{\emph{Phys. Rev. D}
  {\bfseries 59} (1999) 083505}
  [\href{https://arxiv.org/abs/astro-ph/9807098}{{\ttfamily
  astro-ph/9807098}}].

\bibitem{Fuller:1993sp}
G.M.~Fuller, K.~Jedamzik, G.J.~Mathews and A.~Olinto, \emph{{On Constraining
  electroweak baryogenesis with primordial nucleosynthesis}},
  \href{https://doi.org/10.1016/0370-2693(94)91019-7}{\emph{Phys. Lett. B}
  {\bfseries 333} (1994) 135}
  [\href{https://arxiv.org/abs/astro-ph/9407034}{{\ttfamily
  astro-ph/9407034}}].

\bibitem{Heckler:1994uu}
A.F.~Heckler, \emph{{The Effects of electroweak phase transition dynamics on
  baryogenesis and primordial nucleosynthesis}},
  \href{https://doi.org/10.1103/PhysRevD.51.405}{\emph{Phys. Rev. D} {\bfseries
  51} (1995) 405} [\href{https://arxiv.org/abs/astro-ph/9407064}{{\ttfamily
  astro-ph/9407064}}].

\bibitem{Megevand:2004ry}
A.~Megevand and F.~Astorga, \emph{{Generation of baryon inhomogeneities in the
  electroweak phase transition}},
  \href{https://doi.org/10.1103/PhysRevD.71.023502}{\emph{Phys. Rev. D}
  {\bfseries 71} (2005) 023502}
  [\href{https://arxiv.org/abs/hep-ph/0409321}{{\ttfamily hep-ph/0409321}}].

\bibitem{Brandenberger:1994fe}
R.H.~Brandenberger, A.-C.~Davis and M.J.~Rees, \emph{{Nucleosynthesis
  constraints on defect mediated electroweak baryogenesis}},
  \href{https://doi.org/10.1016/0370-2693(95)00272-M}{\emph{Phys. Lett. B}
  {\bfseries 349} (1995) 329}
  [\href{https://arxiv.org/abs/astro-ph/9501040}{{\ttfamily
  astro-ph/9501040}}].

\bibitem{WMAP:2003ivt}
{\scshape WMAP} collaboration, \emph{{First year Wilkinson Microwave Anisotropy
  Probe (WMAP) observations: Preliminary maps and basic results}},
  \href{https://doi.org/10.1086/377253}{\emph{Astrophys. J. Suppl.} {\bfseries
  148} (2003) 1} [\href{https://arxiv.org/abs/astro-ph/0302207}{{\ttfamily
  astro-ph/0302207}}].

\bibitem{Barrow:2018yyg}
J.D.~Barrow and R.J.~Scherrer, \emph{{Constraining Density Fluctuations with
  Big Bang Nucleosynthesis in the Era of Precision Cosmology}},
  \href{https://doi.org/10.1103/PhysRevD.98.043534}{\emph{Phys. Rev. D}
  {\bfseries 98} (2018) 043534}
  [\href{https://arxiv.org/abs/1803.02383}{{\ttfamily 1803.02383}}].

\bibitem{Inomata:2018htm}
K.~Inomata, M.~Kawasaki, A.~Kusenko and L.~Yang, \emph{{Big Bang
  Nucleosynthesis Constraint on Baryonic Isocurvature Perturbations}},
  \href{https://doi.org/10.1088/1475-7516/2018/12/003}{\emph{JCAP} {\bfseries
  12} (2018) 003} [\href{https://arxiv.org/abs/1806.00123}{{\ttfamily
  1806.00123}}].

\bibitem{Burns:2023sgx}
A.-K.~Burns, T.M.P.~Tait and M.~Valli, \emph{{PRyMordial: the first three
  minutes, within and beyond the standard model}},
  \href{https://doi.org/10.1140/epjc/s10052-024-12442-0}{\emph{Eur. Phys. J. C}
  {\bfseries 84} (2024) 86} [\href{https://arxiv.org/abs/2307.07061}{{\ttfamily
  2307.07061}}].

\bibitem{Mukhanov:2003xs}
V.F.~Mukhanov, \emph{{Nucleosynthesis without a computer}},
  \href{https://doi.org/10.1023/B:IJTP.0000048169.69609.77}{\emph{Int. J.
  Theor. Phys.} {\bfseries 43} (2004) 669}
  [\href{https://arxiv.org/abs/astro-ph/0303073}{{\ttfamily
  astro-ph/0303073}}].

\bibitem{Pitaevskii1981Physical}
L.P.~Pitaevskii and E.M.~Lifshitz, \emph{Physical Kinetics: Volume 10 (Course
  of Theoretical Physics)}, Butterworth-Heinemann (Jan., 1981).

\bibitem{AlbornozVasquez:2012emy}
D.~Albornoz~Vasquez, A.~Belikov, A.~Coc, J.~Silk and E.~Vangioni,
  \emph{{Neutron injection during primordial nucleosynthesis alleviates the
  primordial 7Li problem}},
  \href{https://doi.org/10.1103/PhysRevD.86.063501}{\emph{Phys. Rev. D}
  {\bfseries 86} (2012) 063501}
  [\href{https://arxiv.org/abs/1208.0443}{{\ttfamily 1208.0443}}].

\bibitem{Coc:2013eha}
A.~Coc, J.-P.~Uzan and E.~Vangioni, \emph{{Mirror matter can alleviate the
  cosmological lithium problem}},
  \href{https://doi.org/10.1103/PhysRevD.87.123530}{\emph{Phys. Rev. D}
  {\bfseries 87} (2013) 123530}
  [\href{https://arxiv.org/abs/1303.1935}{{\ttfamily 1303.1935}}].

\bibitem{Coc:2014gia}
A.~Coc, M.~Pospelov, J.-P.~Uzan and E.~Vangioni, \emph{{Modified big bang
  nucleosynthesis with nonstandard neutron sources}},
  \href{https://doi.org/10.1103/PhysRevD.90.085018}{\emph{Phys. Rev. D}
  {\bfseries 90} (2014) 085018}
  [\href{https://arxiv.org/abs/1405.1718}{{\ttfamily 1405.1718}}].

\bibitem{Cooke:2017cwo}
R.J.~Cooke, M.~Pettini and C.C.~Steidel, \emph{{One Percent Determination of
  the Primordial Deuterium Abundance}},
  \href{https://doi.org/10.3847/1538-4357/aaab53}{\emph{Astrophys. J.}
  {\bfseries 855} (2018) 102}
  [\href{https://arxiv.org/abs/1710.11129}{{\ttfamily 1710.11129}}].

\bibitem{Zavarygin:2017cov}
E.O.~Zavarygin, J.K.~Webb, V.~Dumont and S.~Riemer-S\o{}rensen, \emph{{The
  primordial deuterium abundance at zabs~=~2.504 from a high signal-to-noise
  spectrum of Q1009+2956}},
  \href{https://doi.org/10.1093/mnras/sty1003}{\emph{Mon. Not. Roy. Astron.
  Soc.} {\bfseries 477} (2018) 5536}
  [\href{https://arxiv.org/abs/1706.09512}{{\ttfamily 1706.09512}}].

\bibitem{SimonsObservatory:2018koc}
{\scshape Simons Observatory} collaboration, \emph{{The Simons Observatory:
  Science goals and forecasts}},
  \href{https://doi.org/10.1088/1475-7516/2019/02/056}{\emph{JCAP} {\bfseries
  02} (2019) 056} [\href{https://arxiv.org/abs/1808.07445}{{\ttfamily
  1808.07445}}].

\bibitem{CMB-S4:2016ple}
{\scshape CMB-S4} collaboration, \emph{{CMB-S4 Science Book, First Edition}},
  \href{https://arxiv.org/abs/1610.02743}{{\ttfamily 1610.02743}}.

\bibitem{Cooke:2024nqz}
R.~Cooke, \emph{{Big Bang Nucleosynthesis}},
  \href{https://arxiv.org/abs/2409.06015}{{\ttfamily 2409.06015}}.

\bibitem{Yeh:2020mgl}
T.-H.~Yeh, K.A.~Olive and B.D.~Fields, \emph{{The impact of new $d(p,\gamma)$3
  rates on Big Bang Nucleosynthesis}},
  \href{https://doi.org/10.1088/1475-7516/2021/03/046}{\emph{JCAP} {\bfseries
  03} (2021) 046} [\href{https://arxiv.org/abs/2011.13874}{{\ttfamily
  2011.13874}}].

\bibitem{talkgustavino}
C.~Gustavino, \emph{{The Baryon density of the Universe from an improved rate
  of deuterium burning}}, {\emph{LA THUILE 2025 - Les Rencontres de Physique de
  la Vallée d'Aoste, 2025} }.

\bibitem{Turner:1992tz}
M.S.~Turner, E.J.~Weinberg and L.M.~Widrow, \emph{{Bubble nucleation in first
  order inflation and other cosmological phase transitions}},
  \href{https://doi.org/10.1103/PhysRevD.46.2384}{\emph{Phys. Rev. D}
  {\bfseries 46} (1992) 2384}.

\bibitem{Futurework1}
H.~Bagherian, M.~Ekhterachian and S.~Stelzl, \emph{{Inhomogeneities as a probe
  of electroweak baryogenesis, work in progress}}, .

\bibitem{LISA:2017pwj}
{\scshape LISA} collaboration, \emph{{Laser Interferometer Space Antenna}},
  \href{https://arxiv.org/abs/1702.00786}{{\ttfamily 1702.00786}}.

\bibitem{Xu:2023wog}
H.~Xu et~al., \emph{{Searching for the Nano-Hertz Stochastic Gravitational Wave
  Background with the Chinese Pulsar Timing Array Data Release I}},
  \href{https://doi.org/10.1088/1674-4527/acdfa5}{\emph{Res. Astron.
  Astrophys.} {\bfseries 23} (2023) 075024}
  [\href{https://arxiv.org/abs/2306.16216}{{\ttfamily 2306.16216}}].

\bibitem{NANOGrav:2023gor}
{\scshape NANOGrav} collaboration, \emph{{The NANOGrav 15 yr Data Set: Evidence
  for a Gravitational-wave Background}},
  \href{https://doi.org/10.3847/2041-8213/acdac6}{\emph{Astrophys. J. Lett.}
  {\bfseries 951} (2023) L8}
  [\href{https://arxiv.org/abs/2306.16213}{{\ttfamily 2306.16213}}].

\bibitem{EPTA:2023fyk}
{\scshape EPTA, InPTA:} collaboration, \emph{{The second data release from the
  European Pulsar Timing Array - III. Search for gravitational wave signals}},
  \href{https://doi.org/10.1051/0004-6361/202346844}{\emph{Astron. Astrophys.}
  {\bfseries 678} (2023) A50}
  [\href{https://arxiv.org/abs/2306.16214}{{\ttfamily 2306.16214}}].

\bibitem{Reardon:2023gzh}
D.J.~Reardon et~al., \emph{{Search for an Isotropic Gravitational-wave
  Background with the Parkes Pulsar Timing Array}},
  \href{https://doi.org/10.3847/2041-8213/acdd02}{\emph{Astrophys. J. Lett.}
  {\bfseries 951} (2023) L6}
  [\href{https://arxiv.org/abs/2306.16215}{{\ttfamily 2306.16215}}].

\bibitem{Caprini:2024hue}
{\scshape LISA Cosmology Working Group} collaboration, \emph{{Gravitational
  waves from first-order phase transitions in LISA: reconstruction pipeline and
  physics interpretation}},
  \href{https://doi.org/10.1088/1475-7516/2024/10/020}{\emph{JCAP} {\bfseries
  10} (2024) 020} [\href{https://arxiv.org/abs/2403.03723}{{\ttfamily
  2403.03723}}].

\bibitem{NANOGrav:2023hvm}
{\scshape NANOGrav} collaboration, \emph{{The NANOGrav 15 yr Data Set: Search
  for Signals from New Physics}},
  \href{https://doi.org/10.3847/2041-8213/acdc91}{\emph{Astrophys. J. Lett.}
  {\bfseries 951} (2023) L11}
  [\href{https://arxiv.org/abs/2306.16219}{{\ttfamily 2306.16219}}].

\bibitem{ACT:2025tim}
{\scshape ACT} collaboration, \emph{{The Atacama Cosmology Telescope: DR6
  Constraints on Extended Cosmological Models}},
  \href{https://arxiv.org/abs/2503.14454}{{\ttfamily 2503.14454}}.

\bibitem{Ferreira:2024eru}
R.Z.~Ferreira, A.~Notari, O.~Pujol\`as and F.~Rompineve, \emph{{Collapsing
  domain wall networks: impact on pulsar timing arrays and primordial black
  holes}}, \href{https://doi.org/10.1088/1475-7516/2024/06/020}{\emph{JCAP}
  {\bfseries 06} (2024) 020}
  [\href{https://arxiv.org/abs/2401.14331}{{\ttfamily 2401.14331}}].

\end{thebibliography}\endgroup

\end{document}